\documentclass[a4paper,UKenglish,cleveref,autoref]{lipics-v2019}
\nolinenumbers
\usepackage{microtype}

\usepackage{booktabs} %
\usepackage{outlines}

\usepackage{listings}
\usepackage{xcolor}

\usepackage[numbers]{natbib}

\definecolor{codegreen}{rgb}{0,0.6,0}
\definecolor{codegray}{rgb}{0.5,0.5,0.5}
\definecolor{codepurple}{rgb}{0.58,0,0.82}
\definecolor{codered}{rgb}{0.82,0.15,0.23}
\definecolor{backcolour}{rgb}{0.95,0.95,0.92}

\lstdefinestyle{mystyle}{
    language=SAS,
    breakatwhitespace=true,
    breaklines=true,
    captionpos=b,
    keepspaces=true,
    numbers=left,
    numbersep=5pt,
    showspaces=false,
    showstringspaces=false,
    showtabs=false,
    tabsize=2,
    aboveskip=20pt,
    belowskip=10pt,
    xleftmargin=0.5cm,
    basicstyle=\footnotesize\ttfamily,
    commentstyle=\itshape\color{codegray},
    numberstyle=\footnotesize\color{codegray},
    stringstyle=\color{codegreen},
    keywordstyle = {\color{codepurple}},
    keywordstyle = [2]{\color{codered}},
    keywordstyle = [3]{\color{codered}},
    keywordstyle = [4]{\color{codegreen}},
    keywords={method,object,interface,type,var,def,return,class,for,in,if},
    otherkeywords = {:,->},
    morekeywords = [2]{:},
    morekeywords = [3]{->},
    morekeywords = [4]{true,false},
    morestring=*[d]{"},
    backgroundcolor={}
}

\def\SOMns{SOM{\sc ns}\xspace}
\def\AWFY{Are\,We\,Fast\,Yet\xspace}

\newcommand{\ie}{i.e.\xspace}
\newcommand{\eg}{e.g.\xspace}

\newcommand{\code}[1]{\texttt{#1}}

\usepackage{letltxmacro}
\LetLtxMacro{\OrgCitep}{\citep}
\renewcommand{\citep}[1]{\,\OrgCitep{#1}}

\usepackage{ifthen}
\newcommand{\citeurl}[5]{%
#1\footnote{\emph{#2}%
          \ifthenelse{\equal{#3}{}}%
                     {}%
                     {, #3}%
          \ifthenelse{\equal{#4}{}}%
                     {}%
                     {, access date: #4}%
, \url{#5}}}

\makeatletter
\@ifpackageloaded{tex4ht}
  {}%
  {\usepackage{tikz}
\tikzstyle{every picture}+=[font=\sffamily]}
\makeatother

\definecolor{fgcolor}{rgb}{0.345, 0.345, 0.345}
\definecolor{messagecolor}{rgb}{0, 0, 0}
\definecolor{errorcolor}{rgb}{1, 0, 0}

\newenvironment{knitrout}{}{} %

\usepackage{framed}
\makeatletter
 {\par\unskip\endMakeFramed%
 \at@end@of@kframe}
\makeatother
\usepackage{alltt}

\def\TypingOverhead{%
\begin{knitrout}
\definecolor{shadecolor}{rgb}{0.969, 0.969, 0.969}

\end{knitrout}
}%

\newcommand{\OverheadTypingGMeanP}{6\%\xspace}
\newcommand{\OverheadTypingMinP}{-14\%\xspace}
\newcommand{\OverheadTypingMaxP}{76\%\xspace}

\newcommand{\OverheadListP}{76\%\xspace}

\newcommand{\OverheadPermuteP}{-14\%\xspace}
\newcommand{\OverheadGraphSearchP}{-4\%\xspace}
\newcommand{\OverheadStorageP}{-5\%\xspace}

\newcommand{\OverheadRichardsP}{38\%\xspace}
\newcommand{\OverheadCDP}{15\%\xspace}
\newcommand{\OverheadSnakeP}{13\%\xspace}
\newcommand{\OverheadTowersP}{12\%\xspace}

\def\TypingStatsTable{%
\begin{tabular}{llccc}
\hline
Type Test & Enabled Optimization & \multicolumn{1}{r}{mean \#invocations} & \multicolumn{1}{r}{min} & \multicolumn{1}{r}{max} \\ 
\hline
check\_generic & Neither  & \multicolumn{1}{r}{137,525,845} & \multicolumn{1}{r}{11,628,068} & \multicolumn{1}{r}{896,604,537} \\
 & Subtype Cache  & \multicolumn{1}{r}{137,525,845} & \multicolumn{1}{r}{11,628,068} & \multicolumn{1}{r}{896,604,537} \\
 & Optimized Node  & \multicolumn{1}{r}{292} & \multicolumn{1}{r}{68} & \multicolumn{1}{r}{1,012} \\
 & Both  & \multicolumn{1}{r}{292} & \multicolumn{1}{r}{68} & \multicolumn{1}{r}{1,012} \\
is\_subtype\_of & Neither  & \multicolumn{1}{r}{134,125,215} & \multicolumn{1}{r}{11,628,067} & \multicolumn{1}{r}{896,604,534} \\
 & Subtype Cache  & \multicolumn{1}{r}{16} & \multicolumn{1}{r}{10} & \multicolumn{1}{r}{29} \\
 & Optimized Node  & \multicolumn{1}{r}{292} & \multicolumn{1}{r}{68} & \multicolumn{1}{r}{1,012} \\
 & Both  & \multicolumn{1}{r}{16} & \multicolumn{1}{r}{10} & \multicolumn{1}{r}{29} \\
\hline 
\end{tabular}

}%

\def\AwfyBaseline{%
\begin{knitrout}
\definecolor{shadecolor}{rgb}{0.969, 0.969, 0.969}
\begin{tikzpicture}[x=1pt,y=1pt]
\definecolor{fillColor}{RGB}{255,255,255}
\path[use as bounding box,fill=fillColor,fill opacity=0.00] (0,0) rectangle (325.21,180.67);
\begin{scope}
\path[clip] ( 62.92, 42.11) rectangle (325.21,180.67);
\definecolor{drawColor}{gray}{0.80}

\path[draw=drawColor,line width= 0.6pt,dash pattern=on 4pt off 4pt ,line join=round] ( 94.14, 42.11) -- ( 94.14,180.67);

\path[draw=drawColor,line width= 0.6pt,dash pattern=on 4pt off 4pt ,line join=round] (107.83, 42.11) -- (107.83,180.67);

\path[draw=drawColor,line width= 0.6pt,dash pattern=on 4pt off 4pt ,line join=round] (140.82, 42.11) -- (140.82,180.67);

\path[draw=drawColor,line width= 0.6pt,dash pattern=on 4pt off 4pt ,line join=round] (160.11, 42.11) -- (160.11,180.67);

\path[draw=drawColor,line width= 0.6pt,dash pattern=on 4pt off 4pt ,line join=round] (173.80, 42.11) -- (173.80,180.67);
\definecolor{drawColor}{gray}{0.20}

\path[draw=drawColor,line width= 0.6pt,line join=round] (251.09, 61.90) -- (325.21, 61.90);

\path[draw=drawColor,line width= 0.6pt,line join=round] (179.71, 61.90) -- (127.70, 61.90);
\definecolor{fillColor}{RGB}{255,255,255}

\path[draw=drawColor,line width= 0.6pt,line join=round,line cap=round,fill=fillColor] (251.09, 49.53) --
	(179.71, 49.53) --
	(179.71, 74.27) --
	(251.09, 74.27) --
	(251.09, 49.53) --
	cycle;

\path[draw=drawColor,line width= 1.1pt,line join=round] (198.42, 49.53) -- (198.42, 74.27);
\definecolor{fillColor}{gray}{0.20}

\path[draw=drawColor,line width= 0.4pt,line join=round,line cap=round,fill=fillColor] (107.83, 94.89) circle (  0.89);

\path[draw=drawColor,line width= 0.6pt,line join=round] (107.83, 94.89) -- (107.83, 94.89);

\path[draw=drawColor,line width= 0.6pt,line join=round] (107.83, 94.89) -- (107.83, 94.89);
\definecolor{fillColor}{RGB}{255,255,255}

\path[draw=drawColor,line width= 0.6pt,line join=round,line cap=round,fill=fillColor] (107.83, 82.52) --
	(107.83, 82.52) --
	(107.83,107.27) --
	(107.83,107.27) --
	(107.83, 82.52) --
	cycle;

\path[draw=drawColor,line width= 1.1pt,line join=round] (107.83, 82.52) -- (107.83,107.27);

\path[draw=drawColor,line width= 0.6pt,line join=round] (169.72,127.89) -- (177.61,127.89);

\path[draw=drawColor,line width= 0.6pt,line join=round] (132.50,127.89) -- (104.61,127.89);

\path[draw=drawColor,line width= 0.6pt,line join=round,line cap=round,fill=fillColor] (169.72,115.51) --
	(132.50,115.51) --
	(132.50,140.26) --
	(169.72,140.26) --
	(169.72,115.51) --
	cycle;

\path[draw=drawColor,line width= 1.1pt,line join=round] (151.54,115.51) -- (151.54,140.26);
\definecolor{fillColor}{gray}{0.20}

\path[draw=drawColor,line width= 0.4pt,line join=round,line cap=round,fill=fillColor] (157.94,160.88) circle (  0.89);

\path[draw=drawColor,line width= 0.4pt,line join=round,line cap=round,fill=fillColor] (112.38,160.88) circle (  0.89);

\path[draw=drawColor,line width= 0.4pt,line join=round,line cap=round,fill=fillColor] ( 98.72,160.88) circle (  0.89);

\path[draw=drawColor,line width= 0.6pt,line join=round] (141.06,160.88) -- (154.23,160.88);

\path[draw=drawColor,line width= 0.6pt,line join=round] (129.86,160.88) -- (115.96,160.88);
\definecolor{fillColor}{RGB}{255,255,255}

\path[draw=drawColor,line width= 0.6pt,line join=round,line cap=round,fill=fillColor] (141.06,148.51) --
	(129.86,148.51) --
	(129.86,173.25) --
	(141.06,173.25) --
	(141.06,148.51) --
	cycle;

\path[draw=drawColor,line width= 1.1pt,line join=round] (137.61,148.51) -- (137.61,173.25);
\end{scope}
\begin{scope}
\path[clip] (  0.00,  0.00) rectangle (325.21,180.67);
\definecolor{drawColor}{RGB}{190,190,190}

\path[draw=drawColor,line width= 0.6pt,line join=round] ( 62.92, 42.11) --
	( 62.92,180.67);
\end{scope}
\begin{scope}
\path[clip] (  0.00,  0.00) rectangle (325.21,180.67);
\definecolor{drawColor}{gray}{0.30}

\node[text=drawColor,anchor=base east,inner sep=0pt, outer sep=0pt, scale=  0.80] at ( 57.97, 59.15) {Higgs};

\node[text=drawColor,anchor=base east,inner sep=0pt, outer sep=0pt, scale=  0.80] at ( 57.97, 92.14) {Java};

\node[text=drawColor,anchor=base east,inner sep=0pt, outer sep=0pt, scale=  0.80] at ( 57.97,125.13) {Moth};

\node[text=drawColor,anchor=base east,inner sep=0pt, outer sep=0pt, scale=  0.80] at ( 57.97,158.12) {Node.js (V8)};
\end{scope}
\begin{scope}
\path[clip] (  0.00,  0.00) rectangle (325.21,180.67);
\definecolor{drawColor}{gray}{0.20}

\path[draw=drawColor,line width= 0.6pt,line join=round] ( 60.17, 61.90) --
	( 62.92, 61.90);

\path[draw=drawColor,line width= 0.6pt,line join=round] ( 60.17, 94.89) --
	( 62.92, 94.89);

\path[draw=drawColor,line width= 0.6pt,line join=round] ( 60.17,127.89) --
	( 62.92,127.89);

\path[draw=drawColor,line width= 0.6pt,line join=round] ( 60.17,160.88) --
	( 62.92,160.88);
\end{scope}
\begin{scope}
\path[clip] (  0.00,  0.00) rectangle (325.21,180.67);
\definecolor{drawColor}{RGB}{190,190,190}

\path[draw=drawColor,line width= 0.6pt,line join=round] ( 62.92, 42.11) --
	(325.21, 42.11);
\end{scope}
\begin{scope}
\path[clip] (  0.00,  0.00) rectangle (325.21,180.67);
\definecolor{drawColor}{gray}{0.20}

\path[draw=drawColor,line width= 0.6pt,line join=round] ( 94.14, 39.36) --
	( 94.14, 42.11);

\path[draw=drawColor,line width= 0.6pt,line join=round] (107.83, 39.36) --
	(107.83, 42.11);

\path[draw=drawColor,line width= 0.6pt,line join=round] (140.82, 39.36) --
	(140.82, 42.11);

\path[draw=drawColor,line width= 0.6pt,line join=round] (160.11, 39.36) --
	(160.11, 42.11);

\path[draw=drawColor,line width= 0.6pt,line join=round] (173.80, 39.36) --
	(173.80, 42.11);

\path[draw=drawColor,line width= 0.6pt,line join=round] (217.41, 39.36) --
	(217.41, 42.11);

\path[draw=drawColor,line width= 0.6pt,line join=round] (294.00, 39.36) --
	(294.00, 42.11);
\end{scope}
\begin{scope}
\path[clip] (  0.00,  0.00) rectangle (325.21,180.67);
\definecolor{drawColor}{gray}{0.30}

\node[text=drawColor,rotate= 90.00,anchor=base east,inner sep=0pt, outer sep=0pt, scale=  0.80] at ( 96.89, 37.16) {0.75};

\node[text=drawColor,rotate= 90.00,anchor=base east,inner sep=0pt, outer sep=0pt, scale=  0.80] at (110.58, 37.16) {1.00};

\node[text=drawColor,rotate= 90.00,anchor=base east,inner sep=0pt, outer sep=0pt, scale=  0.80] at (143.57, 37.16) {2.00};

\node[text=drawColor,rotate= 90.00,anchor=base east,inner sep=0pt, outer sep=0pt, scale=  0.80] at (162.87, 37.16) {3.00};

\node[text=drawColor,rotate= 90.00,anchor=base east,inner sep=0pt, outer sep=0pt, scale=  0.80] at (176.56, 37.16) {4.00};

\node[text=drawColor,rotate= 90.00,anchor=base east,inner sep=0pt, outer sep=0pt, scale=  0.80] at (220.16, 37.16) {10.00};

\node[text=drawColor,rotate= 90.00,anchor=base east,inner sep=0pt, outer sep=0pt, scale=  0.80] at (296.75, 37.16) {50.00};
\end{scope}
\begin{scope}
\path[clip] (  0.00,  0.00) rectangle (325.21,180.67);
\definecolor{drawColor}{RGB}{0,0,0}

\node[text=drawColor,anchor=base,inner sep=0pt, outer sep=0pt, scale=  0.80] at (194.07, 12.46) {Run-time factor, normalized to Java (untyped)};

\node[text=drawColor,anchor=base,inner sep=0pt, outer sep=0pt, scale=  0.80] at (194.07,  3.82) {(lower is better)};
\end{scope}
\begin{scope}
\path[clip] (  0.00,  0.00) rectangle (325.21,180.67);
\definecolor{drawColor}{RGB}{0,0,0}

\node[text=drawColor,rotate= 90.00,anchor=base,inner sep=0pt, outer sep=0pt, scale=  0.80] at (  8.36,111.39) {VM};
\end{scope}
\end{tikzpicture}
 
\end{knitrout}
}%

\newcommand{\OverheadNodeGMeanX}{1.8x\xspace}
\newcommand{\OverheadNodeMinX}{0.8x\xspace}
\newcommand{\OverheadNodeMaxX}{2.9x\xspace}

\newcommand{\OverheadMothGMeanX}{2.3x\xspace}
\newcommand{\OverheadMothMinX}{0.9x\xspace}
\newcommand{\OverheadMothMaxX}{4.3x\xspace}

\newcommand{\OverheadMothNodeMaxX}{2.3x\xspace}

\newcommand{\OverheadMothNodeGMeanP}{32\%\xspace}
\newcommand{\OverheadMothNodeMinP}{-18\%\xspace}

\newcommand{\OverheadHiggsGMeanX}{10.6x\xspace}
\newcommand{\OverheadHiggsMinX}{1.5x\xspace}
\newcommand{\OverheadHiggsMaxX}{169x\xspace}

\newcommand{\WarmupCutOff}{350\xspace}

\def\OptimizationOverview{%
\begin{knitrout}
\definecolor{shadecolor}{rgb}{0.969, 0.969, 0.969}
\begin{tikzpicture}[x=1pt,y=1pt]
\definecolor{fillColor}{RGB}{255,255,255}
\path[use as bounding box,fill=fillColor,fill opacity=0.00] (0,0) rectangle (325.21,180.67);
\begin{scope}
\path[clip] ( 99.31, 46.11) rectangle (325.21,180.67);
\definecolor{drawColor}{gray}{0.80}

\path[draw=drawColor,line width= 0.6pt,dash pattern=on 4pt off 4pt ,line join=round] (128.69, 46.11) -- (128.69,180.67);

\path[draw=drawColor,line width= 0.6pt,dash pattern=on 4pt off 4pt ,line join=round] (134.54, 46.11) -- (134.54,180.67);

\path[draw=drawColor,line width= 0.6pt,dash pattern=on 4pt off 4pt ,line join=round] (159.49, 46.11) -- (159.49,180.67);

\path[draw=drawColor,line width= 0.6pt,dash pattern=on 4pt off 4pt ,line join=round] (209.41, 46.11) -- (209.41,180.67);

\path[draw=drawColor,line width= 0.6pt,dash pattern=on 4pt off 4pt ,line join=round] (257.00, 46.11) -- (257.00,180.67);
\definecolor{drawColor}{gray}{0.20}

\path[draw=drawColor,line width= 0.6pt,line join=round] (286.98, 61.63) -- (314.23, 61.63);

\path[draw=drawColor,line width= 0.6pt,line join=round] (250.62, 61.63) -- (212.72, 61.63);
\definecolor{fillColor}{RGB}{255,255,255}

\path[draw=drawColor,line width= 0.6pt,line join=round,line cap=round,fill=fillColor] (286.98, 51.93) --
	(250.62, 51.93) --
	(250.62, 71.34) --
	(286.98, 71.34) --
	(286.98, 51.93) --
	cycle;

\path[draw=drawColor,line width= 1.1pt,line join=round] (257.56, 51.93) -- (257.56, 71.34);

\path[draw=drawColor,line width= 0.6pt,line join=round] (223.08, 87.51) -- (252.22, 87.51);

\path[draw=drawColor,line width= 0.6pt,line join=round] (189.70, 87.51) -- (166.09, 87.51);

\path[draw=drawColor,line width= 0.6pt,line join=round,line cap=round,fill=fillColor] (223.08, 77.81) --
	(189.70, 77.81) --
	(189.70, 97.22) --
	(223.08, 97.22) --
	(223.08, 77.81) --
	cycle;

\path[draw=drawColor,line width= 1.1pt,line join=round] (207.44, 77.81) -- (207.44, 97.22);
\definecolor{fillColor}{gray}{0.20}

\path[draw=drawColor,line width= 0.4pt,line join=round,line cap=round,fill=fillColor] (155.74,113.39) circle (  0.89);

\path[draw=drawColor,line width= 0.4pt,line join=round,line cap=round,fill=fillColor] (146.30,113.39) circle (  0.89);

\path[draw=drawColor,line width= 0.6pt,line join=round] (136.95,113.39) -- (138.89,113.39);

\path[draw=drawColor,line width= 0.6pt,line join=round] (133.63,113.39) -- (129.78,113.39);
\definecolor{fillColor}{RGB}{255,255,255}

\path[draw=drawColor,line width= 0.6pt,line join=round,line cap=round,fill=fillColor] (136.95,103.69) --
	(133.63,103.69) --
	(133.63,123.09) --
	(136.95,123.09) --
	(136.95,103.69) --
	cycle;

\path[draw=drawColor,line width= 1.1pt,line join=round] (134.46,103.69) -- (134.46,123.09);
\definecolor{fillColor}{gray}{0.20}

\path[draw=drawColor,line width= 0.4pt,line join=round,line cap=round,fill=fillColor] (154.97,139.27) circle (  0.89);

\path[draw=drawColor,line width= 0.6pt,line join=round] (138.64,139.27) -- (146.20,139.27);

\path[draw=drawColor,line width= 0.6pt,line join=round] (133.54,139.27) -- (129.02,139.27);
\definecolor{fillColor}{RGB}{255,255,255}

\path[draw=drawColor,line width= 0.6pt,line join=round,line cap=round,fill=fillColor] (138.64,129.56) --
	(133.54,129.56) --
	(133.54,148.97) --
	(138.64,148.97) --
	(138.64,129.56) --
	cycle;

\path[draw=drawColor,line width= 1.1pt,line join=round] (134.72,129.56) -- (134.72,148.97);
\definecolor{fillColor}{gray}{0.20}

\path[draw=drawColor,line width= 0.4pt,line join=round,line cap=round,fill=fillColor] (134.54,165.15) circle (  0.89);

\path[draw=drawColor,line width= 0.4pt,line join=round,line cap=round,fill=fillColor] (134.54,165.15) circle (  0.89);

\path[draw=drawColor,line width= 0.4pt,line join=round,line cap=round,fill=fillColor] (134.54,165.15) circle (  0.89);

\path[draw=drawColor,line width= 0.4pt,line join=round,line cap=round,fill=fillColor] (134.54,165.15) circle (  0.89);

\path[draw=drawColor,line width= 0.6pt,line join=round] (134.54,165.15) -- (134.54,165.15);

\path[draw=drawColor,line width= 0.6pt,line join=round] (134.54,165.15) -- (134.54,165.15);
\definecolor{fillColor}{RGB}{255,255,255}

\path[draw=drawColor,line width= 0.6pt,line join=round,line cap=round,fill=fillColor] (134.54,155.44) --
	(134.54,155.44) --
	(134.54,174.85) --
	(134.54,174.85) --
	(134.54,155.44) --
	cycle;

\path[draw=drawColor,line width= 1.1pt,line join=round] (134.54,155.44) -- (134.54,174.85);
\end{scope}
\begin{scope}
\path[clip] (  0.00,  0.00) rectangle (325.21,180.67);
\definecolor{drawColor}{RGB}{190,190,190}

\path[draw=drawColor,line width= 0.6pt,line join=round] ( 99.31, 46.11) --
	( 99.31,180.67);
\end{scope}
\begin{scope}
\path[clip] (  0.00,  0.00) rectangle (325.21,180.67);
\definecolor{drawColor}{gray}{0.30}

\node[text=drawColor,anchor=base east,inner sep=0pt, outer sep=0pt, scale=  0.80] at ( 94.36, 58.88) {Moth (neither)};

\node[text=drawColor,anchor=base east,inner sep=0pt, outer sep=0pt, scale=  0.80] at ( 94.36, 84.76) {Moth (subtype cache)};

\node[text=drawColor,anchor=base east,inner sep=0pt, outer sep=0pt, scale=  0.80] at ( 94.36,110.64) {Moth (optimized node)};

\node[text=drawColor,anchor=base east,inner sep=0pt, outer sep=0pt, scale=  0.80] at ( 94.36,136.51) {Moth (both)};

\node[text=drawColor,anchor=base east,inner sep=0pt, outer sep=0pt, scale=  0.80] at ( 94.36,162.39) {Moth (untyped)};
\end{scope}
\begin{scope}
\path[clip] (  0.00,  0.00) rectangle (325.21,180.67);
\definecolor{drawColor}{gray}{0.20}

\path[draw=drawColor,line width= 0.6pt,line join=round] ( 96.56, 61.63) --
	( 99.31, 61.63);

\path[draw=drawColor,line width= 0.6pt,line join=round] ( 96.56, 87.51) --
	( 99.31, 87.51);

\path[draw=drawColor,line width= 0.6pt,line join=round] ( 96.56,113.39) --
	( 99.31,113.39);

\path[draw=drawColor,line width= 0.6pt,line join=round] ( 96.56,139.27) --
	( 99.31,139.27);

\path[draw=drawColor,line width= 0.6pt,line join=round] ( 96.56,165.15) --
	( 99.31,165.15);
\end{scope}
\begin{scope}
\path[clip] (  0.00,  0.00) rectangle (325.21,180.67);
\definecolor{drawColor}{RGB}{190,190,190}

\path[draw=drawColor,line width= 0.6pt,line join=round] ( 99.31, 46.11) --
	(325.21, 46.11);
\end{scope}
\begin{scope}
\path[clip] (  0.00,  0.00) rectangle (325.21,180.67);
\definecolor{drawColor}{gray}{0.20}

\path[draw=drawColor,line width= 0.6pt,line join=round] (128.69, 43.36) --
	(128.69, 46.11);

\path[draw=drawColor,line width= 0.6pt,line join=round] (134.54, 43.36) --
	(134.54, 46.11);

\path[draw=drawColor,line width= 0.6pt,line join=round] (159.49, 43.36) --
	(159.49, 46.11);

\path[draw=drawColor,line width= 0.6pt,line join=round] (209.41, 43.36) --
	(209.41, 46.11);

\path[draw=drawColor,line width= 0.6pt,line join=round] (257.00, 43.36) --
	(257.00, 46.11);

\path[draw=drawColor,line width= 0.6pt,line join=round] (275.39, 43.36) --
	(275.39, 46.11);

\path[draw=drawColor,line width= 0.6pt,line join=round] (300.35, 43.36) --
	(300.35, 46.11);

\path[draw=drawColor,line width= 0.6pt,line join=round] (314.95, 43.36) --
	(314.95, 46.11);
\end{scope}
\begin{scope}
\path[clip] (  0.00,  0.00) rectangle (325.21,180.67);
\definecolor{drawColor}{gray}{0.30}

\node[text=drawColor,rotate= 90.00,anchor=base east,inner sep=0pt, outer sep=0pt, scale=  0.80] at (131.44, 41.16) {0.85};

\node[text=drawColor,rotate= 90.00,anchor=base east,inner sep=0pt, outer sep=0pt, scale=  0.80] at (137.29, 41.16) {1.00};

\node[text=drawColor,rotate= 90.00,anchor=base east,inner sep=0pt, outer sep=0pt, scale=  0.80] at (162.25, 41.16) {2.00};

\node[text=drawColor,rotate= 90.00,anchor=base east,inner sep=0pt, outer sep=0pt, scale=  0.80] at (212.16, 41.16) {8.00};

\node[text=drawColor,rotate= 90.00,anchor=base east,inner sep=0pt, outer sep=0pt, scale=  0.80] at (259.75, 41.16) {30.00};

\node[text=drawColor,rotate= 90.00,anchor=base east,inner sep=0pt, outer sep=0pt, scale=  0.80] at (278.15, 41.16) {50.00};

\node[text=drawColor,rotate= 90.00,anchor=base east,inner sep=0pt, outer sep=0pt, scale=  0.80] at (303.10, 41.16) {100.00};

\node[text=drawColor,rotate= 90.00,anchor=base east,inner sep=0pt, outer sep=0pt, scale=  0.80] at (317.70, 41.16) {150.00};
\end{scope}
\begin{scope}
\path[clip] (  0.00,  0.00) rectangle (325.21,180.67);
\definecolor{drawColor}{RGB}{0,0,0}

\node[text=drawColor,anchor=base,inner sep=0pt, outer sep=0pt, scale=  0.80] at (212.26, 12.46) {Run-time factor, normalized to Moth (untyped)};

\node[text=drawColor,anchor=base,inner sep=0pt, outer sep=0pt, scale=  0.80] at (212.26,  3.82) {(lower is better)};
\end{scope}
\begin{scope}
\path[clip] (  0.00,  0.00) rectangle (325.21,180.67);
\definecolor{drawColor}{RGB}{0,0,0}

\node[text=drawColor,rotate= 90.00,anchor=base,inner sep=0pt, outer sep=0pt, scale=  0.80] at (  8.36,113.39) {VM};
\end{scope}
\end{tikzpicture}
 
\end{knitrout}
}%

\def\TypeCostFirstIt{%
\begin{knitrout}
\definecolor{shadecolor}{rgb}{0.969, 0.969, 0.969}
\begin{tikzpicture}[x=1pt,y=1pt]
\definecolor{fillColor}{RGB}{255,255,255}
\path[use as bounding box,fill=fillColor,fill opacity=0.00] (0,0) rectangle (187.90,180.67);
\begin{scope}
\path[clip] ( 42.64, 22.70) rectangle (112.52,164.42);
\definecolor{fillColor}{RGB}{248,118,109}

\path[fill=fillColor,fill opacity=0.30] ( 45.81, 47.89) --
	( 58.52, 44.72) --
	( 71.22, 44.20) --
	( 83.93, 49.01) --
	( 96.64, 44.66) --
	(109.34, 50.72) --
	(109.34, 41.38) --
	( 96.64, 34.62) --
	( 83.93, 40.50) --
	( 71.22, 34.12) --
	( 58.52, 34.04) --
	( 45.81, 37.20) --
	cycle;
\definecolor{fillColor}{RGB}{0,186,56}

\path[fill=fillColor,fill opacity=0.30] ( 45.81, 40.33) --
	( 58.52, 81.16) --
	( 71.22, 96.26) --
	( 83.93,109.18) --
	( 96.64,127.27) --
	(109.34,135.82) --
	(109.34,126.66) --
	( 96.64,115.83) --
	( 83.93, 97.42) --
	( 71.22, 87.18) --
	( 58.52, 72.23) --
	( 45.81, 29.98) --
	cycle;
\definecolor{fillColor}{RGB}{97,156,255}

\path[fill=fillColor,fill opacity=0.30] ( 45.81, 42.82) --
	( 58.52, 44.15) --
	( 71.22, 47.46) --
	( 83.93, 37.56) --
	( 96.64, 40.37) --
	(109.34, 39.98) --
	(109.34, 30.01) --
	( 96.64, 29.84) --
	( 83.93, 29.14) --
	( 71.22, 36.76) --
	( 58.52, 32.72) --
	( 45.81, 31.91) --
	cycle;
\definecolor{drawColor}{RGB}{248,118,109}

\path[draw=drawColor,line width= 0.6pt,line join=round] ( 45.81, 42.55) --
	( 58.52, 39.38) --
	( 71.22, 39.16) --
	( 83.93, 44.76) --
	( 96.64, 39.64) --
	(109.34, 46.05);
\definecolor{drawColor}{RGB}{0,186,56}

\path[draw=drawColor,line width= 0.6pt,line join=round] ( 45.81, 35.15) --
	( 58.52, 76.69) --
	( 71.22, 91.72) --
	( 83.93,103.30) --
	( 96.64,121.55) --
	(109.34,131.24);
\definecolor{drawColor}{RGB}{97,156,255}

\path[draw=drawColor,line width= 0.6pt,line join=round] ( 45.81, 37.37) --
	( 58.52, 38.44) --
	( 71.22, 42.11) --
	( 83.93, 33.35) --
	( 96.64, 35.11) --
	(109.34, 34.99);
\end{scope}
\begin{scope}
\path[clip] (118.02, 22.70) rectangle (187.90,164.42);
\definecolor{fillColor}{RGB}{248,118,109}

\path[fill=fillColor,fill opacity=0.30] (121.20, 60.95) --
	(133.90, 72.13) --
	(146.61, 68.99) --
	(159.31, 71.75) --
	(172.02, 68.88) --
	(184.73, 76.88) --
	(184.73, 69.36) --
	(172.02, 61.99) --
	(159.31, 64.10) --
	(146.61, 59.70) --
	(133.90, 62.30) --
	(121.20, 50.59) --
	cycle;
\definecolor{fillColor}{RGB}{0,186,56}

\path[fill=fillColor,fill opacity=0.30] (121.20, 64.85) --
	(133.90, 98.95) --
	(146.61,115.06) --
	(159.31,130.59) --
	(172.02,141.39) --
	(184.73,157.98) --
	(184.73,150.32) --
	(172.02,131.89) --
	(159.31,120.01) --
	(146.61,108.04) --
	(133.90, 91.09) --
	(121.20, 53.44) --
	cycle;
\definecolor{fillColor}{RGB}{97,156,255}

\path[fill=fillColor,fill opacity=0.30] (121.20, 55.11) --
	(133.90, 56.15) --
	(146.61, 59.84) --
	(159.31, 60.42) --
	(172.02, 60.93) --
	(184.73, 58.78) --
	(184.73, 49.72) --
	(172.02, 53.45) --
	(159.31, 49.56) --
	(146.61, 50.76) --
	(133.90, 48.34) --
	(121.20, 47.24) --
	cycle;
\definecolor{drawColor}{RGB}{248,118,109}

\path[draw=drawColor,line width= 0.6pt,line join=round] (121.20, 55.77) --
	(133.90, 67.22) --
	(146.61, 64.35) --
	(159.31, 67.93) --
	(172.02, 65.43) --
	(184.73, 73.12);
\definecolor{drawColor}{RGB}{0,186,56}

\path[draw=drawColor,line width= 0.6pt,line join=round] (121.20, 59.15) --
	(133.90, 95.02) --
	(146.61,111.55) --
	(159.31,125.30) --
	(172.02,136.64) --
	(184.73,154.15);
\definecolor{drawColor}{RGB}{97,156,255}

\path[draw=drawColor,line width= 0.6pt,line join=round] (121.20, 51.18) --
	(133.90, 52.24) --
	(146.61, 55.30) --
	(159.31, 54.99) --
	(172.02, 57.19) --
	(184.73, 54.25);
\end{scope}
\begin{scope}
\path[clip] ( 42.64,164.42) rectangle (112.52,180.67);
\definecolor{drawColor}{gray}{0.10}

\node[text=drawColor,anchor=base,inner sep=0pt, outer sep=0pt, scale=  0.80] at ( 77.58,169.79) {Check};
\end{scope}
\begin{scope}
\path[clip] (118.02,164.42) rectangle (187.90,180.67);
\definecolor{drawColor}{gray}{0.10}

\node[text=drawColor,anchor=base,inner sep=0pt, outer sep=0pt, scale=  0.80] at (152.96,169.79) {Nest};
\end{scope}
\begin{scope}
\path[clip] (  0.00,  0.00) rectangle (187.90,180.67);
\definecolor{drawColor}{RGB}{190,190,190}

\path[draw=drawColor,line width= 0.6pt,line join=round] ( 42.64, 22.70) --
	(112.52, 22.70);
\end{scope}
\begin{scope}
\path[clip] (  0.00,  0.00) rectangle (187.90,180.67);
\definecolor{drawColor}{gray}{0.20}

\path[draw=drawColor,line width= 0.6pt,line join=round] ( 45.81, 19.95) --
	( 45.81, 22.70);

\path[draw=drawColor,line width= 0.6pt,line join=round] ( 58.52, 19.95) --
	( 58.52, 22.70);

\path[draw=drawColor,line width= 0.6pt,line join=round] ( 71.22, 19.95) --
	( 71.22, 22.70);

\path[draw=drawColor,line width= 0.6pt,line join=round] ( 83.93, 19.95) --
	( 83.93, 22.70);

\path[draw=drawColor,line width= 0.6pt,line join=round] ( 96.64, 19.95) --
	( 96.64, 22.70);

\path[draw=drawColor,line width= 0.6pt,line join=round] (109.34, 19.95) --
	(109.34, 22.70);
\end{scope}
\begin{scope}
\path[clip] (  0.00,  0.00) rectangle (187.90,180.67);
\definecolor{drawColor}{gray}{0.30}

\node[text=drawColor,anchor=base,inner sep=0pt, outer sep=0pt, scale=  0.80] at ( 45.81, 12.24) {0};

\node[text=drawColor,anchor=base,inner sep=0pt, outer sep=0pt, scale=  0.80] at ( 58.52, 12.24) {1};

\node[text=drawColor,anchor=base,inner sep=0pt, outer sep=0pt, scale=  0.80] at ( 71.22, 12.24) {2};

\node[text=drawColor,anchor=base,inner sep=0pt, outer sep=0pt, scale=  0.80] at ( 83.93, 12.24) {3};

\node[text=drawColor,anchor=base,inner sep=0pt, outer sep=0pt, scale=  0.80] at ( 96.64, 12.24) {4};

\node[text=drawColor,anchor=base,inner sep=0pt, outer sep=0pt, scale=  0.80] at (109.34, 12.24) {5};
\end{scope}
\begin{scope}
\path[clip] (  0.00,  0.00) rectangle (187.90,180.67);
\definecolor{drawColor}{RGB}{190,190,190}

\path[draw=drawColor,line width= 0.6pt,line join=round] (118.02, 22.70) --
	(187.90, 22.70);
\end{scope}
\begin{scope}
\path[clip] (  0.00,  0.00) rectangle (187.90,180.67);
\definecolor{drawColor}{gray}{0.20}

\path[draw=drawColor,line width= 0.6pt,line join=round] (121.20, 19.95) --
	(121.20, 22.70);

\path[draw=drawColor,line width= 0.6pt,line join=round] (133.90, 19.95) --
	(133.90, 22.70);

\path[draw=drawColor,line width= 0.6pt,line join=round] (146.61, 19.95) --
	(146.61, 22.70);

\path[draw=drawColor,line width= 0.6pt,line join=round] (159.31, 19.95) --
	(159.31, 22.70);

\path[draw=drawColor,line width= 0.6pt,line join=round] (172.02, 19.95) --
	(172.02, 22.70);

\path[draw=drawColor,line width= 0.6pt,line join=round] (184.73, 19.95) --
	(184.73, 22.70);
\end{scope}
\begin{scope}
\path[clip] (  0.00,  0.00) rectangle (187.90,180.67);
\definecolor{drawColor}{gray}{0.30}

\node[text=drawColor,anchor=base,inner sep=0pt, outer sep=0pt, scale=  0.80] at (121.20, 12.24) {0};

\node[text=drawColor,anchor=base,inner sep=0pt, outer sep=0pt, scale=  0.80] at (133.90, 12.24) {1};

\node[text=drawColor,anchor=base,inner sep=0pt, outer sep=0pt, scale=  0.80] at (146.61, 12.24) {2};

\node[text=drawColor,anchor=base,inner sep=0pt, outer sep=0pt, scale=  0.80] at (159.31, 12.24) {3};

\node[text=drawColor,anchor=base,inner sep=0pt, outer sep=0pt, scale=  0.80] at (172.02, 12.24) {4};

\node[text=drawColor,anchor=base,inner sep=0pt, outer sep=0pt, scale=  0.80] at (184.73, 12.24) {5};
\end{scope}
\begin{scope}
\path[clip] (  0.00,  0.00) rectangle (187.90,180.67);
\definecolor{drawColor}{RGB}{190,190,190}

\path[draw=drawColor,line width= 0.6pt,line join=round] ( 42.64, 22.70) --
	( 42.64,164.42);
\end{scope}
\begin{scope}
\path[clip] (  0.00,  0.00) rectangle (187.90,180.67);
\definecolor{drawColor}{gray}{0.30}

\node[text=drawColor,anchor=base east,inner sep=0pt, outer sep=0pt, scale=  0.80] at ( 37.69, 26.99) {2100};

\node[text=drawColor,anchor=base east,inner sep=0pt, outer sep=0pt, scale=  0.80] at ( 37.69, 59.99) {2400};

\node[text=drawColor,anchor=base east,inner sep=0pt, outer sep=0pt, scale=  0.80] at ( 37.69, 92.99) {2700};

\node[text=drawColor,anchor=base east,inner sep=0pt, outer sep=0pt, scale=  0.80] at ( 37.69,125.99) {3000};

\node[text=drawColor,anchor=base east,inner sep=0pt, outer sep=0pt, scale=  0.80] at ( 37.69,158.99) {3300};
\end{scope}
\begin{scope}
\path[clip] (  0.00,  0.00) rectangle (187.90,180.67);
\definecolor{drawColor}{gray}{0.20}

\path[draw=drawColor,line width= 0.6pt,line join=round] ( 39.89, 29.74) --
	( 42.64, 29.74);

\path[draw=drawColor,line width= 0.6pt,line join=round] ( 39.89, 62.74) --
	( 42.64, 62.74);

\path[draw=drawColor,line width= 0.6pt,line join=round] ( 39.89, 95.74) --
	( 42.64, 95.74);

\path[draw=drawColor,line width= 0.6pt,line join=round] ( 39.89,128.74) --
	( 42.64,128.74);

\path[draw=drawColor,line width= 0.6pt,line join=round] ( 39.89,161.74) --
	( 42.64,161.74);
\end{scope}
\begin{scope}
\path[clip] (  0.00,  0.00) rectangle (187.90,180.67);
\definecolor{drawColor}{RGB}{0,0,0}

\node[text=drawColor,anchor=base,inner sep=0pt, outer sep=0pt, scale=  0.80] at (115.27,  3.82) {Number of Typed Method Arguments};
\end{scope}
\begin{scope}
\path[clip] (  0.00,  0.00) rectangle (187.90,180.67);
\definecolor{drawColor}{RGB}{0,0,0}

\node[text=drawColor,rotate= 90.00,anchor=base,inner sep=0pt, outer sep=0pt, scale=  0.80] at (  8.36, 93.56) {Run time (ms)};

\node[text=drawColor,rotate= 90.00,anchor=base,inner sep=0pt, outer sep=0pt, scale=  0.80] at ( 17.00, 93.56) {(lower is better)};
\end{scope}
\end{tikzpicture}
 
\end{knitrout}
}%

\def\TypeCostLastIt{%
\begin{knitrout}
\definecolor{shadecolor}{rgb}{0.969, 0.969, 0.969}
\begin{tikzpicture}[x=1pt,y=1pt]
\definecolor{fillColor}{RGB}{255,255,255}
\path[use as bounding box,fill=fillColor,fill opacity=0.00] (0,0) rectangle (187.90,180.67);
\begin{scope}
\path[clip] ( 38.64, 22.70) rectangle (110.52,164.42);
\definecolor{fillColor}{RGB}{248,118,109}

\path[fill=fillColor,fill opacity=0.30] ( 41.90, 32.61) --
	( 54.97, 31.53) --
	( 68.04, 32.78) --
	( 81.11, 32.49) --
	( 94.18, 33.47) --
	(107.25, 35.17) --
	(107.25, 31.20) --
	( 94.18, 30.26) --
	( 81.11, 30.34) --
	( 68.04, 30.30) --
	( 54.97, 29.81) --
	( 41.90, 29.93) --
	cycle;
\definecolor{fillColor}{RGB}{0,186,56}

\path[fill=fillColor,fill opacity=0.30] ( 41.90, 33.50) --
	( 54.97, 61.55) --
	( 68.04, 85.15) --
	( 81.11,110.64) --
	( 94.18,134.26) --
	(107.25,157.78) --
	(107.25,150.06) --
	( 94.18,125.76) --
	( 81.11,102.04) --
	( 68.04, 79.91) --
	( 54.97, 57.36) --
	( 41.90, 30.03) --
	cycle;
\definecolor{fillColor}{RGB}{97,156,255}

\path[fill=fillColor,fill opacity=0.30] ( 41.90, 32.55) --
	( 54.97, 31.41) --
	( 68.04, 32.86) --
	( 81.11, 32.85) --
	( 94.18, 34.46) --
	(107.25, 33.21) --
	(107.25, 30.25) --
	( 94.18, 30.75) --
	( 81.11, 30.25) --
	( 68.04, 30.38) --
	( 54.97, 29.90) --
	( 41.90, 29.98) --
	cycle;
\definecolor{drawColor}{RGB}{248,118,109}

\path[draw=drawColor,line width= 0.6pt,line join=round] ( 41.90, 31.27) --
	( 54.97, 30.67) --
	( 68.04, 31.54) --
	( 81.11, 31.41) --
	( 94.18, 31.86) --
	(107.25, 33.19);
\definecolor{drawColor}{RGB}{0,186,56}

\path[draw=drawColor,line width= 0.6pt,line join=round] ( 41.90, 31.76) --
	( 54.97, 59.45) --
	( 68.04, 82.53) --
	( 81.11,106.34) --
	( 94.18,130.01) --
	(107.25,153.92);
\definecolor{drawColor}{RGB}{97,156,255}

\path[draw=drawColor,line width= 0.6pt,line join=round] ( 41.90, 31.27) --
	( 54.97, 30.65) --
	( 68.04, 31.62) --
	( 81.11, 31.55) --
	( 94.18, 32.60) --
	(107.25, 31.73);
\end{scope}
\begin{scope}
\path[clip] (116.02, 22.70) rectangle (187.90,164.42);
\definecolor{fillColor}{RGB}{248,118,109}

\path[fill=fillColor,fill opacity=0.30] (119.29, 34.29) --
	(132.36, 30.15) --
	(145.43, 33.59) --
	(158.50, 34.09) --
	(171.56, 33.35) --
	(184.63, 35.20) --
	(184.63, 31.03) --
	(171.56, 30.32) --
	(158.50, 30.76) --
	(145.43, 30.27) --
	(132.36, 29.14) --
	(119.29, 31.03) --
	cycle;
\definecolor{fillColor}{RGB}{0,186,56}

\path[fill=fillColor,fill opacity=0.30] (119.29, 35.56) --
	(132.36, 64.04) --
	(145.43, 88.49) --
	(158.50,107.22) --
	(171.56,132.12) --
	(184.63,157.98) --
	(184.63,147.55) --
	(171.56,123.13) --
	(158.50, 98.68) --
	(145.43, 81.81) --
	(132.36, 58.18) --
	(119.29, 30.73) --
	cycle;
\definecolor{fillColor}{RGB}{97,156,255}

\path[fill=fillColor,fill opacity=0.30] (119.29, 33.21) --
	(132.36, 33.53) --
	(145.43, 34.43) --
	(158.50, 34.69) --
	(171.56, 33.22) --
	(184.63, 31.27) --
	(184.63, 29.59) --
	(171.56, 30.50) --
	(158.50, 31.25) --
	(145.43, 30.74) --
	(132.36, 30.82) --
	(119.29, 29.95) --
	cycle;
\definecolor{drawColor}{RGB}{248,118,109}

\path[draw=drawColor,line width= 0.6pt,line join=round] (119.29, 32.66) --
	(132.36, 29.65) --
	(145.43, 31.93) --
	(158.50, 32.42) --
	(171.56, 31.83) --
	(184.63, 33.12);
\definecolor{drawColor}{RGB}{0,186,56}

\path[draw=drawColor,line width= 0.6pt,line join=round] (119.29, 33.14) --
	(132.36, 61.11) --
	(145.43, 85.15) --
	(158.50,102.95) --
	(171.56,127.63) --
	(184.63,152.77);
\definecolor{drawColor}{RGB}{97,156,255}

\path[draw=drawColor,line width= 0.6pt,line join=round] (119.29, 31.58) --
	(132.36, 32.17) --
	(145.43, 32.59) --
	(158.50, 32.97) --
	(171.56, 31.86) --
	(184.63, 30.43);
\end{scope}
\begin{scope}
\path[clip] ( 38.64,164.42) rectangle (110.52,180.67);
\definecolor{drawColor}{gray}{0.10}

\node[text=drawColor,anchor=base,inner sep=0pt, outer sep=0pt, scale=  0.80] at ( 74.58,169.79) {Check};
\end{scope}
\begin{scope}
\path[clip] (116.02,164.42) rectangle (187.90,180.67);
\definecolor{drawColor}{gray}{0.10}

\node[text=drawColor,anchor=base,inner sep=0pt, outer sep=0pt, scale=  0.80] at (151.96,169.79) {Nest};
\end{scope}
\begin{scope}
\path[clip] (  0.00,  0.00) rectangle (187.90,180.67);
\definecolor{drawColor}{RGB}{190,190,190}

\path[draw=drawColor,line width= 0.6pt,line join=round] ( 38.64, 22.70) --
	(110.52, 22.70);
\end{scope}
\begin{scope}
\path[clip] (  0.00,  0.00) rectangle (187.90,180.67);
\definecolor{drawColor}{gray}{0.20}

\path[draw=drawColor,line width= 0.6pt,line join=round] ( 41.90, 19.95) --
	( 41.90, 22.70);

\path[draw=drawColor,line width= 0.6pt,line join=round] ( 54.97, 19.95) --
	( 54.97, 22.70);

\path[draw=drawColor,line width= 0.6pt,line join=round] ( 68.04, 19.95) --
	( 68.04, 22.70);

\path[draw=drawColor,line width= 0.6pt,line join=round] ( 81.11, 19.95) --
	( 81.11, 22.70);

\path[draw=drawColor,line width= 0.6pt,line join=round] ( 94.18, 19.95) --
	( 94.18, 22.70);

\path[draw=drawColor,line width= 0.6pt,line join=round] (107.25, 19.95) --
	(107.25, 22.70);
\end{scope}
\begin{scope}
\path[clip] (  0.00,  0.00) rectangle (187.90,180.67);
\definecolor{drawColor}{gray}{0.30}

\node[text=drawColor,anchor=base,inner sep=0pt, outer sep=0pt, scale=  0.80] at ( 41.90, 12.24) {0};

\node[text=drawColor,anchor=base,inner sep=0pt, outer sep=0pt, scale=  0.80] at ( 54.97, 12.24) {1};

\node[text=drawColor,anchor=base,inner sep=0pt, outer sep=0pt, scale=  0.80] at ( 68.04, 12.24) {2};

\node[text=drawColor,anchor=base,inner sep=0pt, outer sep=0pt, scale=  0.80] at ( 81.11, 12.24) {3};

\node[text=drawColor,anchor=base,inner sep=0pt, outer sep=0pt, scale=  0.80] at ( 94.18, 12.24) {4};

\node[text=drawColor,anchor=base,inner sep=0pt, outer sep=0pt, scale=  0.80] at (107.25, 12.24) {5};
\end{scope}
\begin{scope}
\path[clip] (  0.00,  0.00) rectangle (187.90,180.67);
\definecolor{drawColor}{RGB}{190,190,190}

\path[draw=drawColor,line width= 0.6pt,line join=round] (116.02, 22.70) --
	(187.90, 22.70);
\end{scope}
\begin{scope}
\path[clip] (  0.00,  0.00) rectangle (187.90,180.67);
\definecolor{drawColor}{gray}{0.20}

\path[draw=drawColor,line width= 0.6pt,line join=round] (119.29, 19.95) --
	(119.29, 22.70);

\path[draw=drawColor,line width= 0.6pt,line join=round] (132.36, 19.95) --
	(132.36, 22.70);

\path[draw=drawColor,line width= 0.6pt,line join=round] (145.43, 19.95) --
	(145.43, 22.70);

\path[draw=drawColor,line width= 0.6pt,line join=round] (158.50, 19.95) --
	(158.50, 22.70);

\path[draw=drawColor,line width= 0.6pt,line join=round] (171.56, 19.95) --
	(171.56, 22.70);

\path[draw=drawColor,line width= 0.6pt,line join=round] (184.63, 19.95) --
	(184.63, 22.70);
\end{scope}
\begin{scope}
\path[clip] (  0.00,  0.00) rectangle (187.90,180.67);
\definecolor{drawColor}{gray}{0.30}

\node[text=drawColor,anchor=base,inner sep=0pt, outer sep=0pt, scale=  0.80] at (119.29, 12.24) {0};

\node[text=drawColor,anchor=base,inner sep=0pt, outer sep=0pt, scale=  0.80] at (132.36, 12.24) {1};

\node[text=drawColor,anchor=base,inner sep=0pt, outer sep=0pt, scale=  0.80] at (145.43, 12.24) {2};

\node[text=drawColor,anchor=base,inner sep=0pt, outer sep=0pt, scale=  0.80] at (158.50, 12.24) {3};

\node[text=drawColor,anchor=base,inner sep=0pt, outer sep=0pt, scale=  0.80] at (171.56, 12.24) {4};

\node[text=drawColor,anchor=base,inner sep=0pt, outer sep=0pt, scale=  0.80] at (184.63, 12.24) {5};
\end{scope}
\begin{scope}
\path[clip] (  0.00,  0.00) rectangle (187.90,180.67);
\definecolor{drawColor}{RGB}{190,190,190}

\path[draw=drawColor,line width= 0.6pt,line join=round] ( 38.64, 22.70) --
	( 38.64,164.42);
\end{scope}
\begin{scope}
\path[clip] (  0.00,  0.00) rectangle (187.90,180.67);
\definecolor{drawColor}{gray}{0.30}

\node[text=drawColor,anchor=base east,inner sep=0pt, outer sep=0pt, scale=  0.80] at ( 33.69, 57.27) {200};

\node[text=drawColor,anchor=base east,inner sep=0pt, outer sep=0pt, scale=  0.80] at ( 33.69, 95.85) {400};

\node[text=drawColor,anchor=base east,inner sep=0pt, outer sep=0pt, scale=  0.80] at ( 33.69,134.42) {600};
\end{scope}
\begin{scope}
\path[clip] (  0.00,  0.00) rectangle (187.90,180.67);
\definecolor{drawColor}{gray}{0.20}

\path[draw=drawColor,line width= 0.6pt,line join=round] ( 35.89, 60.03) --
	( 38.64, 60.03);

\path[draw=drawColor,line width= 0.6pt,line join=round] ( 35.89, 98.60) --
	( 38.64, 98.60);

\path[draw=drawColor,line width= 0.6pt,line join=round] ( 35.89,137.17) --
	( 38.64,137.17);
\end{scope}
\begin{scope}
\path[clip] (  0.00,  0.00) rectangle (187.90,180.67);
\definecolor{drawColor}{RGB}{0,0,0}

\node[text=drawColor,anchor=base,inner sep=0pt, outer sep=0pt, scale=  0.80] at (113.27,  3.82) {Number of Typed Method Arguments};
\end{scope}
\begin{scope}
\path[clip] (  0.00,  0.00) rectangle (187.90,180.67);
\definecolor{drawColor}{RGB}{0,0,0}

\node[text=drawColor,rotate= 90.00,anchor=base,inner sep=0pt, outer sep=0pt, scale=  0.80] at (  8.36, 93.56) {Run time (ms)};

\node[text=drawColor,rotate= 90.00,anchor=base,inner sep=0pt, outer sep=0pt, scale=  0.80] at ( 17.00, 93.56) {(lower is better)};
\end{scope}
\begin{scope}
\path[clip] (  0.00,  0.00) rectangle (187.90,180.67);
\definecolor{fillColor}{RGB}{255,255,255}

\path[fill=fillColor] ( 39.13,154.64) rectangle ( 53.58,169.10);
\end{scope}
\begin{scope}
\path[clip] (  0.00,  0.00) rectangle (187.90,180.67);
\definecolor{fillColor}{RGB}{248,118,109}

\path[fill=fillColor,fill opacity=0.30] ( 39.84,155.35) rectangle ( 52.87,168.38);
\end{scope}
\begin{scope}
\path[clip] (  0.00,  0.00) rectangle (187.90,180.67);
\definecolor{drawColor}{RGB}{248,118,109}

\path[draw=drawColor,line width= 0.6pt,line join=round] ( 40.57,161.87) -- ( 52.14,161.87);
\end{scope}
\begin{scope}
\path[clip] (  0.00,  0.00) rectangle (187.90,180.67);
\definecolor{fillColor}{RGB}{255,255,255}

\path[fill=fillColor] ( 39.13,140.19) rectangle ( 53.58,154.64);
\end{scope}
\begin{scope}
\path[clip] (  0.00,  0.00) rectangle (187.90,180.67);
\definecolor{fillColor}{RGB}{0,186,56}

\path[fill=fillColor,fill opacity=0.30] ( 39.84,140.90) rectangle ( 52.87,153.93);
\end{scope}
\begin{scope}
\path[clip] (  0.00,  0.00) rectangle (187.90,180.67);
\definecolor{drawColor}{RGB}{0,186,56}

\path[draw=drawColor,line width= 0.6pt,line join=round] ( 40.57,147.42) -- ( 52.14,147.42);
\end{scope}
\begin{scope}
\path[clip] (  0.00,  0.00) rectangle (187.90,180.67);
\definecolor{fillColor}{RGB}{255,255,255}

\path[fill=fillColor] ( 39.13,125.73) rectangle ( 53.58,140.19);
\end{scope}
\begin{scope}
\path[clip] (  0.00,  0.00) rectangle (187.90,180.67);
\definecolor{fillColor}{RGB}{97,156,255}

\path[fill=fillColor,fill opacity=0.30] ( 39.84,126.45) rectangle ( 52.87,139.48);
\end{scope}
\begin{scope}
\path[clip] (  0.00,  0.00) rectangle (187.90,180.67);
\definecolor{drawColor}{RGB}{97,156,255}

\path[draw=drawColor,line width= 0.6pt,line join=round] ( 40.57,132.96) -- ( 52.14,132.96);
\end{scope}
\begin{scope}
\path[clip] (  0.00,  0.00) rectangle (187.90,180.67);
\definecolor{drawColor}{RGB}{0,0,0}

\node[text=drawColor,anchor=base west,inner sep=0pt, outer sep=0pt, scale=  0.80] at ( 53.58,159.11) {Moth (both)};
\end{scope}
\begin{scope}
\path[clip] (  0.00,  0.00) rectangle (187.90,180.67);
\definecolor{drawColor}{RGB}{0,0,0}

\node[text=drawColor,anchor=base west,inner sep=0pt, outer sep=0pt, scale=  0.80] at ( 53.58,144.66) {Moth (neither)};
\end{scope}
\begin{scope}
\path[clip] (  0.00,  0.00) rectangle (187.90,180.67);
\definecolor{drawColor}{RGB}{0,0,0}

\node[text=drawColor,anchor=base west,inner sep=0pt, outer sep=0pt, scale=  0.80] at ( 53.58,130.21) {Moth (untyped)};
\end{scope}
\end{tikzpicture}
 
\end{knitrout}
}%
 
\usepackage[hideall]{collab}
\collabAuthor{rr}{green!60!black}{Richard}
\collabAuthor{sm}{red}{Stefan}
\collabAuthor{kjx}{orange}{James}
\collabAuthor{mwh}{purple}{Michael}

\title{Transient Typechecks are (Almost) Free [Preliminary Draft 2019-02-18]}

\author{Richard Roberts}{School of Design, Victoria University of Wellington}{rykardo.r@gmail.com}{https://orcid.org/0000-0002-3462-8539}{}

\author{Stefan Marr}{School of Computing, University of Kent}{s.marr@kent.ac.uk}{https://orcid.org0000-0001-9059-5180}{}

\author{Michael Homer}{School of Engineering and Computer Science, Victoria University of Wellington}{mwh@ecs.vuw.ac.nz}{}{}

\author{James Noble}{School of Engineering and Computer Science, Victoria University of Wellington}{kjx@ecs.vuw.ac.nz}{https://orcid.org/0000-0001-9036-5692}{}

\authorrunning{Roberts, Marr, Homer, Noble}

\Copyright{Richard Roberts and Stefan Marr and Michael Homer and James Noble}



\funding{This work is supported by the Royal Society of New Zealand
  Marsden Fund}
\EventEditors{John Q. Open and Joan R. Access}
\EventNoEds{2}
\EventLongTitle{42nd Conference on Very Important Topics (CVIT 2016)}
\EventShortTitle{CVIT 2016}
\EventAcronym{CVIT}
\EventYear{2016}
\EventDate{December 24--27, 2016}
\EventLocation{Little Whinging, United Kingdom}
\EventLogo{}
\SeriesVolume{42}
\ArticleNo{23}

\begin{document}

\maketitle

\begin{abstract}
 Transient gradual typing imposes run-time type tests that typically cause a 
 linear slowdown in programs' performance.
 This performance impact discourages the use of type annotations
 because adding types to a program makes the program slower.
 A virtual machine can employ standard just-in-time optimizations to
 reduce the overhead of transient checks to near zero.
 These optimizations can give gradually-typed languages
 performance comparable to state-of-the-art dynamic languages,
 so programmers can add types to their code
 without affecting their programs' performance.
\end{abstract}

\section{Introduction}
\label{sec:introduction}

\begin{verse}
  \textit{``It is a truth universally acknowledged, that a dynamic language
    in possession of a good user base, must be in want of a type system.''}
\end{verse}
\vspace*{-6ex}
\begin{flushright}
with apologies to Jane Austen.
\end{flushright}

Dynamic languages are increasingly prominent in the software industry.
Building on the pioneering work of Self\citep{Self}, 
much work in academia and industry
has gone into making them more efficient\citep{Bolz2013,Bolz:2013:IMT,Wurthinger:2017:PPE,Daloze2016,Clifford:2015:MM,Degenbaev:2016:ITG}.
Just-in-time compilers have, for example, turned JavaScript from a
na{\"\i}vely interpreted language barely suitable for browser scripting, 
into a highly efficient ecosystem, eagerly adopted
by professional programmers for a very wide range of tasks \cite{Pang2018}.

A key advantage of these dynamic languages is the flexibility offered
by the lack of a static type system. 
From the perspective of many computer scientists, software engineers,
and \ugh{computational theologists}\sm{can we drop this?}, this advantage has a concomitant
disadvantage in that programs without types are considered more
difficult to read, to understand, and to analyze than programs with
types. Gradual Typing aims to remedy this disadvantage, adding types
to dynamic languages while maintaining their flexibility
\citep{GiladPluggable2004,Siek2006,XXXSiek2015}. 

There is a spectrum of different approaches to gradual typing
\cite{kafka18,bensurvey18icfp}.
At one end\,---\,``pluggable types'' as in Strongtalk \cite{strongtalk} or ``erasure
semantics'' as in 
TypeScript\citep{typeScriptECOOP}\,---\,
all types are erased before the execution, limiting the benefit of
types to the statically typed parts of programs, and preventing
programs from depending on type checks at run time.  In the middle,
``transient'' or ``type-tag'' checks as in Reticulated Python 
offer first-order semantics, checking
whether an object's type constructor or supported methods match
explicit type declarations
\cite{Siek2007,Bloom2009,concrete15,reticPython2014,Greenman2018}.
Reticulated Python also supports an alternative ``monotonic'' semantics
which mutates an object to narrow its concrete type when it is passed
into a more specific type context.
At the other end of the spectrum, behavioral
typechecks as in Typed Racket \cite{typedScheme08,takikawa2012},
Gradualtalk \cite{gradualtalk14},
and Reticulated Python's proxies,
support higher-order semantics, retaining
types until run 
time, performing the checks eagerly, and giving detailed information
about type violations as soon as possible via blame
tracking \cite{blame2009,blameForAll2011}.
Unfortunately, any gradual system with run-time semantics
(i.e.\ everything more complex than erasure) currently
imposes significant run-time performance overhead to provide those semantics
\citep{Takikawa2016,Vitousek2017,Muehlboeck2017,Bauman2017,Richards2017,Stulova2016,Greenman2018}.

The performance cost of run-time checks is problematic in itself,
but also creates perverse incentives. Rather than the ideal of
gradually adding types in the process of hardening a developing
program, the programmer is incentivized to leave the program untyped
or even to \textit{remove} existing types in search of speed.
While the Gradual Guarantee~\cite{XXXSiek2015} requires that
removing a type annotation does not affect the result of the
program, the performance profile can be drastically shifted by the
overhead of ill-placed checks.
For programs with crucial performance constraints, for new
programmers, and for gradual language designers, juggling this
overhead can lead to increased complexity, suboptimal
software-engineering choices, and code that is harder to maintain,
debug, and analyze.

In this paper, we focus on the centre of the gradual typing
spectrum: the transient, first-order, type-tag checks as used in
Reticulated Python and similar systems. 
Several studies have found
that these type checks have a negative impact on programs'
performance.
Chung, Li, Nardelli and Vitek, for example, found that 
``\textit{The transient approach checks types at uses, so the act of
  adding types to a program introduces more casts and may slow the
  program down (even in fully typed code).}'' and say 
``\textit{"transient semantics\ldots is a worst case scenario\ldots,
  there is a cast at almost every call"}'' \cite{kafka18}.
Greenman and Felleisen find that the slowdown is predictable, as
transient checking ``\textit{imposes a run-time checking overhead that
  is directly proportional to the number of [type annotations] in
  the program"}'' \cite{bensurvey18icfp}, and 
Greenman and Migeed found a ``\textit{clear trend
that adding type annotations adds performance overhead. The increase
is typically linear.}'' \cite{Greenman2018}.

In contrast, we demonstrate that
transient type checks can be ``almost
free'' via a just-in-time compiler to an
optimizing virtual machine.
We insert 
gradual checks
na\"ively, for each gradual type 
annotation.
Whenever an annotated method is called or
returns, or
an annotated variable is accessed,
we check types dynamically, and
terminate the program with a type error if the check fails.
Despite this simplistic approach, a just-in-time compiler can
eliminate redundant checks---%
removing almost all of the checking overhead,
resulting in a performance profile aligned with untyped code.

We evaluate our approach by adding transient type checks to Moth,
an implementation of the Grace programming language
built on top of Truffle
and the Graal just-in-time compiler\citep{Wurthinger2013,Wurthinger:2017:PPE}.
Inspired by Richards~\textit{et~al.}~\cite{Richards2017} and
Bauman~\textit{et~al.}~\cite{Bauman2017},
our approach conflates types
with information about the dynamic object structure 
(maps\citep{Self} or object shapes\citep{woss2014object}), 
which allows the just-in-time compiler
to reduce redundancy between checking structure
and checking types; consequently, most of the overhead
that results from type checking is eliminated.

The contributions of this paper are:

\begin{itemize}
\item demonstrating that VM optimizations enable
        transient gradual type checks with low performance cost
\item an implementation approach that requires
      only small changes to existing abstract-syntax-tree interpreters
\item an evaluation based on classic benchmarks
      and benchmarks from the literature on gradual typing
\end{itemize}

\clearpage

\section{Gradual Types in Grace}
\label{sec:background}

This section introduces Grace, and 
motivates supporting transient gradual typing in the language.

\subsection{The Grace Programming Language}
\label{ssec:grace}

Grace is an object-oriented, imperative, educational programming
language, with a focus on introductory programming
courses, but also intended for more advanced study and research\citep{graceOnward12,graceSigcse13}.
While Grace's syntax draws
from the so-called ``curly bracket'' tradition of C, Java, and
JavaScript, the structure of the language
is in many ways closer to Smalltalk:
all computation is via dynamically dispatched  ``method requests''
where the object receiving the request decides which code to run,
and
returns within lambdas that are ``non-local'', returning to the method
activation in which the block is instantiated\citep{bluebook}.  In
other ways, Grace is closer to JavaScript than Smalltalk: Grace
objects can be created from object literals, rather than by
instantiating classes\citep{Black2007-emeraldHOPL,JonesECOOP2016} and
objects and classes can be deeply nested within each 
other\citep{betabook}.

Critically, Grace's declarations and methods' arguments
and results can be annotated with types, and those types can be  checked
either statically or dynamically. This means the type system is
intrinsically gradual:%
~type annotations should not affect the semantics of a correct
program\citep{XXXSiek2015}, and the type system
includes a distinguished ``\code{Unknown}'' type which matches any other type
and is the implicit type for untyped program parts.

The static core of Grace's type system is well described
elsewhere\citep{TimJonesThesis};%
~here we explain how
these types can be understood 
dynamically, from the Grace programmer's point of view.
Grace's types are structural \citep{graceOnward12},
that is, an object implements a type whenever it
implements all the methods required by that type,
rather than requiring classes or objects to declare types explicitly.
Methods match when they have the same name and arity:
argument and return types are ignored.
A type thus expresses the requests an object can respond to,
for example whether a particular accessor is available,
rather than a nominal location in an explicit inheritance hierarchy.

Grace then checks the types of values at run-time:
\begin{itemize}
\item the values of arguments are checked after a method is requested, 
      but before the body of the method is executed;
\item the value returned by a method is checked after its body is executed; and
\item the values of variables are checked
      whenever written or read by user code.
\end{itemize}
In the spectrum of gradual typing, these semantics are
closest to the
transient typechecks of Reticulated Python
\cite{reticPython2014,Greenman2018}.
Reticulated Python inserts transient checks only when a value flows from untyped
to typed code, while Grace inserts transient checks only at explict type
annotations (but in principle checks every annotation every time).

\subsection{Why Gradual Typing?}

Our primary motivation for this work
is to provide a flexible system 
to check consistency between an execution of a program
and its type annotations.
 A key part of the design philosophy of Grace is that the language should not force
 students to annotate programs with types until they are ready, so that
 teachers can choose whether to introduce types, early, late, or even
 not at all. 

A secondary goal is to have a design that can be implemented with
only a small set of changes to facilitate integration in existing systems.

Both of these goals are shared with
much of the other work on gradual type systems,
but our context leads to some different choices.
First,
while checking Grace's type annotations statically may be optional,
checking them dynamically should not be:%
~any value that flows into a variable, argument, or result
annotated with a type must conform to that type annotation.
Second, 
adding type annotations should not degrade a program's performance,
or rather, programmers should not be encouraged to
improve performance by removing type annotations.
And third, 
we allow the programmer to execute a program even when not statically type-correct.
Allowing such execution is useful to students,
where they can see concrete examples of dynamic type errors.
This is possible because Grace's static type checking is optional,
which means that an
implementation cannot depend on the correctness or mutual
compatibility of a program's type
annotations.

Unfortunately, existing gradual type
implementations do not meet these goals, particularly regarding
performance; hence the ongoing debate about whether gradual typing is
alive, dead, or some state
in between\citep{Takikawa2016,Vitousek2017,Muehlboeck2017,Bauman2017,Richards2017,Greenman2018}.

\subsection{Using Grace's Gradual Types}

We now illustrate how the gradual type checks work in practice
in the context of a simple program to record information about vehicles.
Suppose the programmer starts developing this vehicle
application by defining an object intended to represent a car
(\cref{lst:car-reg}, \cref{ex:object}) and writes a method that, given
the car object, prints out its registration number (\cref{ex:method}).

\begin{lstlisting}[caption={The start of a simple program for tracking vehicle information.},float=h,label=lst:car-reg,escapechar=|,columns=flexible]
def car = object {|\label{ex:object}|
    var registration is public := "JO3553"
}

method printRegistration(v) {|\label{ex:method}|
    print "Registration: {v.registration}"
}
\end{lstlisting}

\begin{lstlisting}[label={ex:vehicle},caption={Adding a type annotation to a method parameter.},escapechar=|,columns=flexible,float,floatplacement=H]
type Vehicle = interface { |\label{ex:adding-type:vehicle}|
    registration    
}

method printRegistration(v: Vehicle) { |\label{ex:adding-type}|
    print "Registration: {v.registration}"
}
\end{lstlisting}

Next, the programmer adds a check to ensure any object passed to the
\code{print\-Reg\-is\-tra\-tion} method will respond to the
\code{registration} request; 
they define the structural type \code{Vehicle}\citep{theCleanVehicle}
naming just that method (\cref{ex:vehicle}, \cref{ex:adding-type:vehicle}), 
and annotate the \code{printRegistration} method's
argument with that type (\cref{ex:vehicle}, \cref{ex:adding-type}).
The annotation ensures that a type error will be thrown if an object,
passed to the \code{printRegistration} method,
cannot respond to the \code{registration} message.
Furthermore, as type errors constituent termination, 
a crash somewhere in the middle of the
implementation of the \code{print} method
will now be avoided.

In \cref{ex:complex}, 
the programmer continues development and creates two car objects 
(\cref{ex:personal-car,ex:government-car}),
that conform to an expanded \code{Vehicle} type (\cref{ex:new-vehicle}).

\begin{lstlisting}[caption={A program in development with inconsistent
    types.},escapechar=|,label={ex:complex},float,floatplacement=htb,columns=flexible,float,floatplacement=H]
type Vehicle = interface { |\label{ex:new-vehicle}|
    registration
    registerTo(_)
}

type Person = interface { name }
type Department = interface { code }

var personalCar : Vehicle := |\label{ex:personal-car}|
  object {
    var registration is public := "DLS018"
    method registerTo(p: Person) {|\label{ex:personal-car:registerTo}|
      print "{p.name} registers {self}"
    } 
  }

var governmentCar : Vehicle := |\label{ex:government-car}|
  object {
    var registration is public := "FKD218"
    method registerTo(d: Department) { |\label{ex:government-car:registerTo}|
      print "some department {self}"
    }
  }

governmentCar.registerTo( |\label{ex:invoke-register-to}|
  object {
    var name is public := "Richard"
  }
)
\end{lstlisting}

Note that each version of the \code{registerTo} method
declares a different type for its parameter
(\cref{ex:personal-car:registerTo,ex:government-car:registerTo}).
When the programmer executes this program,
both \code{personal\-Car} and \code{governmentCar} can be assigned to
a variable declared as \code{Vehicle} because checking that assignment considers only that the vehicle has
a \code{registerTo} method, but not the required argument type of that
method.
At \cref{ex:invoke-register-to} the developer
attempts to register a government car to a person:%
~only when the method is \textit{invoked} (\cref{ex:government-car:registerTo})
will the gradual type test on the argument fail
(the object that is passed in is not a \code{Department} because it lacks a
\code{code} method).

\section{Moth: Grace on Graal and Truffle}
\label{ssec:moth}
\label{sec:method}

Implementing dynamic languages as state-of-the-art virtual machines
can require enormous engineering efforts.
Meta-compilation approaches\citep{Marr:2015:MTPE}
such as RPython\citep{Bolz:2009:TMP,Bolz:2013:IMT}
and GraalVM\citep{Wurthinger2013,Wurthinger:2017:PPE}
reduce the necessary work dramatically,
because they allow language implementers to leverage existing VMs
and their support for just-in-time compilation and garbage collection.

Moth\citep{Roberts2017} adapts \SOMns\citep{SOMns} to leverage this infrastructure for Grace.
\SOMns is a Newspeak implementation\citep{Bracha:10:NS} on top of the Truffle framework and the Graal just-in-time compiler,
which are part of the GraalVM project.
One key optimization of \SOMns for this work is the use of
object shapes\citep{woss2014object},
also known as maps\citep{Self} or hidden classes.
They represent the structure of an object and the types of its fields.
In \SOMns, shapes correspond to the class of an object and augment it with
run-time type information.
With Moth's implementation,
\SOMns was changed to parse Grace code,
adapting a few of the self-optimizing abstract-syntax-tree nodes
to conform to Grace's semantics.
Despite these changes Moth
preserves the peak performance of \SOMns,
which reaches that of V8,
Google's JavaScript implementation
(cf. \cref{sec:baseline-perf} and Marr~\textit{et~al.}~\cite{Marr2016}).

\subsection{Adding Gradual Type Checking} 
\label{ssec:implementation}

One of the goals for our approach to gradual typing was to keep
the necessary changes to an existing implementation small,
while enabling optimization in highly efficient language runtimes.
In an AST interpreter, we can implement this approach by attaching the
checks to the relevant AST nodes: the expected types for the argument
and return values can be included with the node for requesting a
method, and the expected type for a variable can be attached to the
nodes for reading from and writing to that variable.  In practice, we
encapsulate the logic of the check within a new class of AST
nodes, specially to support gradual type checking.  Moth's front end was adapted to parse and record type
annotations and attach instances of this checking node as children of the
existing method, variable read, and variable write nodes.

The check node uses the internal representation of a Grace type
(cf. \cref{ex:type}, \cref{ex:type:check}) to test whether an observed
object conforms to that type. 
An object satisfies a type if all members required by the type are provided
by that object (\cref{ex:type:satisfied}).

\begin{lstlisting}[label={ex:type},escapechar=|,caption={Sketch of a \code{Type} in our system and its \code{check()} semantics.},float,floatplacement=htb,columns=flexible,float,floatplacement=H]
class Type:
  def init(members):
    self._members = members

  def is_satisfied_by(other: Type): |\label{ex:type:satisfied}|
    for m in self._members:
      if m not in other._members:
        return False
    return True

  def check(obj: Object):
    t = get_type(obj)
    return self.is_satisfied_by(t) |\label{ex:type:check}|
\end{lstlisting}

\subsection{Optimization}
\label{ssec:optimization}

There are two aspects to our implementation that are critical for a minimal-overhead solution:

\begin{itemize}
  \item specialized executions of the type checking node, along with guards to protect these specialized versions, and
  \item a matrix to cache sub-typing relationships to eliminate
    redundant exhaustive subtype tests.
\end{itemize}

\begin{lstlisting}[label={ex:typenode},escapechar=|,caption={An illustration of the type checking node that support type checking},float,floatplacement=htbp,columns=flexible,morekeywords={global}]
global record: Matrix

class TypeCheckNode(Node):

  expected: Type

  @Spec(static_guard=expected.check(obj))
  def check(obj: Number): |\label{ex:typenode:number}|
    pass

  @Spec(static_guard=expected.check(obj))
  def check(obj: String): |\label{ex:typenode:string}|
    pass

  ...

  @Spec(
      guard=obj.shape==cached_shape,
      static_guard=expected.check(obj))
  def check(obj: Object, @Cached(obj.shape) cached_shape: Shape): |\label{ex:typenode:object}|
    pass
  
  @Fallback
  def check_generic(obj: Any): |\label{ex:typenode:generic}|
    T = get_type(obj)
    
    if record[T, expected] is unknown: |\label{ex:typenode:matrix}|
      record[T, expected] =
          T.is_subtype_of(expected) |\label{ex:typenode:reuse}|

    if not record[T, expected]:
      raise TypeError(
          "{obj} doesn't implement {expected}")
\end{lstlisting}

The first performance-critical aspect to our implementation
is the optimization of the type checking node.
We rely on Truffle and its TruffleDSL\citep{humer2014domainspecific}.
This means we provide a number of special cases,
which are selected during execution based on the observed concrete 
kinds of objects.
A sketch of our type checking node using a pseudo-code version of the DSL
is given in \cref{ex:typenode}.
A simple optimization is for well known types such as
numbers (\cref{ex:typenode:number}) or strings (\cref{ex:typenode:string}).
The methods annotated with \code{@Spec} (shorthand for \code{@Specialization})
correspond to possible states in a state machine that is generated by the
TruffleDSL.
Thus, if a check node observes a number or a string,
it will check on the first execution only that the expected type,
\ie, the one defined by some type annotation,
is satisfied by the object by using a \code{static\_guard}.
If this is the case, the DSL will activate this state.
For just-in-time compilation, only the activated states and their normal guards are considered.
A \code{static\_guard} is not included in the optimized code.
If a check fails, or no specialization matches, the fallback,
\ie, \code{check\_generic} is selected (\cref{ex:typenode:generic}),
which may raise a type error.

For generic objects, we rely on the specialization on \cref{ex:typenode:object},
which checks that the object satisfies the expected type.
If that is the case, it reads the shape of the object (cf. \cref{ssec:moth}) at specialization time,
and caches it for later comparisons.
Thus, during normal execution,
we only need to read the shape of the object and then compare it to the cached shape
with a simple reference comparison.
If the shapes are the same, we can assume the type check passed successfully.
Note that shapes are not equivalent to types,
however, shapes imply the set of members of an object, and thus,
do imply whether an object fulfills one of our structural types.

The other performance-critical aspect to our implementation
is the use of a matrix to cache sub-typing relationships.
The matrix compares types against types,
featuring all known types along the columns and the same types again along the rows.
A cell in the table corresponds to a sub-typing relationship:
does the type corresponding to the row implement
the type corresponding to the column?
All cells in the matrix begin as unknown and, as
encountered in checks during execution, we populate the table.
If a particular relationship has been computed before
we can skip the check and instead recall the previously-computed value
(\cref{ex:typenode:reuse}).
Using this table we are able to eliminate the redundancy of evaluating
the same type to type relationships across different checks in the program. To reduce redundancy further we also unify types in a similar way to Java's string interning; 
during the construction of a type we first check to see if the same
set of members is expressed by a previously-created type and, if so,
we avoid creating the new instance and provide the existing one instead.

Together the self-specializing type check node and the cache matrix 
ensure that our implementation eliminates redundancy, and
consequently, we are able to minimize the run-time overhead of our system.

\section{Evaluation}
\label{sec:evaluation}

\newcommand{\NumIterationsAll}{1000\xspace}
\newcommand{\NumIterationsHiggs}{100\xspace}

To evaluate our approach to gradual type checking,
we first establish the baseline performance of Moth
compared to Java and JavaScript,
and then assess the impact of the type checks themselves.

\subsection{Method and Setup}

To account for the complex warmup behavior
of modern systems\citep{Barrett:2017:VMW} as well as
the non-determinism caused by \eg garbage collection and cache effects,
we run each benchmark for \NumIterationsAll iterations in the same
VM invocation\footnote{
For the Higgs VM, we only use \NumIterationsHiggs iterations,
because of its lower performance.
This is sufficient since Higgs's compilation approach induces less variation
and leads to more stable measurements.}.
Afterwards, we inspected the run-time plots over the iterations
and manually determined a cutoff of \WarmupCutOff iterations for warmup,
\ie, we discard iterations with signs of compilation.
As a result, we use a large number of data points to compute the average,
but outliers, caused by \eg garbage collection, remain visible in the plots.
All reported averages use the geometric mean since they aggregate ratios.

All experiments were executed on a machine running Ubuntu Linux 16.04.4,
with Kernel 3.13.
The machine has two Intel Xeon E5-2620 v3 2.40GHz,
with 6 cores each, for a total of 24 hyperthreads.
We used ReBench 0.10.1\citep{ReBench:2018}, Java 1.8.0\_171, Graal 0.33 (\code{a13b888}),
Node.js 10.4, and Higgs from 9 May 2018 (\code{aa95240}).
Benchmarks were executed one by one to avoid interference between them.
The analysis of the results was done with R 3.4.1,
and plots are generated with ggplot 2.2.1 and tikzDevice 0.11.
Our experimental setup is available online to enable reproductions.\footnote{
Removed for double blind review
}

\subsection{\AWFY?}
\label{sec:baseline-perf}

\begin{figure}
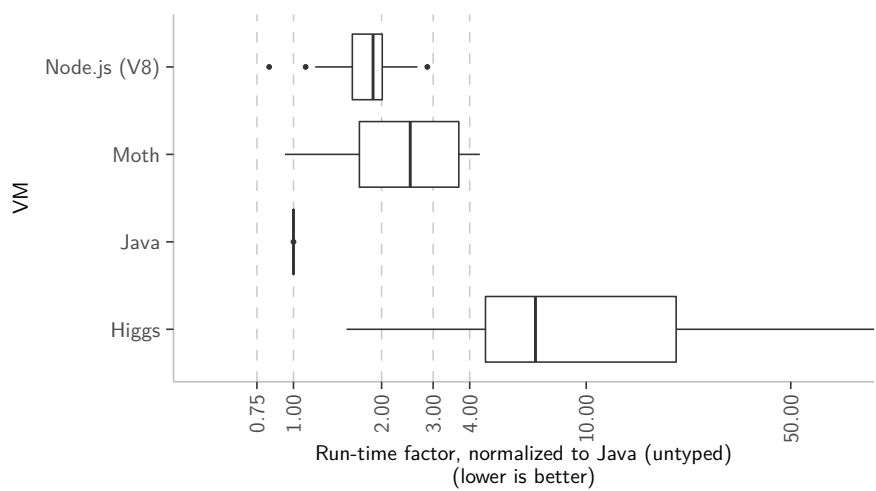

  \centering
	\AwfyBaseline{}
	\caption{Comparison of Java 1.8, Node.js 10.4, Higgs VM, and Moth.
  The boxplot depicts the peak-performance results for the \AWFY benchmarks,
  each benchmark normalized based on the result for Java.
  For these benchmarks, Moth is within the performance range
  of JavaScript, as implemented by Node.js,
  which makes Moth an acceptable platform for our experiments.}
	\label{fig:awfy-baseline}
\end{figure}

To establish the performance of Moth,
we compare it to Java and JavaScript.
For JavaScript we chose two implementations,
Node.js with V8 as well as the Higgs VM.
The Higgs VM is an interesting point of comparison,
because Richards~\textit{et~al.}~\cite{Richards2017} used it in their study.
The goal of this comparison is to determine whether our approach
could be applicable to industrial strength language implementations
without adverse effects on their performance. 

We compare across languages based on the \AWFY benchmarks\citep{Marr2016},
which are designed to enable a comparison
of the effectiveness of compilers across different languages.
To this end, they use only a common set of core language elements.
While this reduces the performance-relevant differences between languages,
the set of core language elements covers only common object-oriented language
features with first-class functions.
Consequently, these benchmarks are not necessarily a predictor
for application performance,
but can give a good indication for basic mechanisms such as type checking.

\Cref{fig:awfy-baseline} shows the results.
We use Java as baseline since it is the fastest language implementation
in this experiment.
We see that Node.js (V8) is about
\OverheadNodeGMeanX (min. \OverheadNodeMinX, max. \OverheadNodeMaxX)
slower than Java.
Moth is about \OverheadMothGMeanX (min. \OverheadMothMinX, max. \OverheadMothMaxX) slower than Java.
As such, it is on average \OverheadMothNodeGMeanP (min. \OverheadMothNodeMinP, max. \OverheadMothNodeMaxX) slower than Node.js.
Compared to the Higgs VM, which is on these benchmarks
\OverheadHiggsGMeanX (min. \OverheadHiggsMinX, max. \OverheadHiggsMaxX) slower than Java,
Moth reaches the performance of Node.js more closely.
With these results, we argue that Moth is a suitable platform to
assess the impact of our approach to gradual type checking,
because its performance is close enough to state-of-the-art VMs,
and run-time overhead is not hidden by slow baseline performance.

\subsection{Performance of Transient Gradual Type Checks}

The performance overhead of our transient gradual type checking system
is assessed based on the \AWFY benchmarks
as well as benchmarks from the gradual-typing literature.
The goal was to complement our benchmarks with additional ones that are
used for similar experiments and can be ported to Grace.
To this end, we surveyed a number of papers\citep{Takikawa2016,Vitousek2017,Muehlboeck2017,Bauman2017,Richards2017,Stulova2016,Greenman2018}
and selected benchmarks that have been used by multiple papers.
Some of these benchmarks overlapped with the \AWFY suite,
or were available in different versions.
While not always behaviorally equivalent,
we chose the \AWFY versions since we already used them to
establish the performance baseline.
The selected benchmarks as well as the papers in which they were used are shown in
\cref{tab:gradual-benchmarks}.

\begin{table}[htb]
  \caption{Benchmarks selected from literature.}
  \label{tab:gradual-benchmarks}
  \begin{center}
    \begin{tabular}{l l r}
      Fannkuch & \cite{Vitousek2017,Greenman2018} \\
      Float & \cite{Vitousek2017,Muehlboeck2017,Greenman2018} \\
      Go & \cite{Vitousek2017,Muehlboeck2017,Greenman2018} \\
      NBody & \cite{Kuhlenschmidt:2018:preprint,Vitousek2017,Greenman2018} & used \cite{Marr2016} \\
      Queens & \cite{Vitousek2017,Muehlboeck2017,Greenman2018} & used \cite{Marr2016} \\
      PyStone & \cite{Vitousek2017,Muehlboeck2017,Greenman2018} \\
      Sieve & \cite{Takikawa2016,Muehlboeck2017,Bauman2017,Richards2017} & used \cite{Marr2016} \\
      Snake & \cite{Takikawa2016,Muehlboeck2017,Bauman2017,Richards2017} \\
      SpectralNorm & \cite{Vitousek2017,Muehlboeck2017,Greenman2018} \\
    \end{tabular}  
  \end{center}
\end{table}

The benchmarks were modified to have complete type information.
To ensure correctness and completeness of these experiments,
we added an additional check to Moth that
reports absent type information to ensure each benchmark is completely typed.
To assess the performance overhead of type checking,
we compare the execution of Moth with all checks disabled, \ie, the baseline version from 
\cref{sec:baseline-perf}, against an execution that has all checks enabled.
We did not measure programs that mix typed and untyped code
because with our implementation technique a fully typed program is expected to
have the largest overhead.

\paragraph*{Peak Performance}

\begin{figure}[htb]
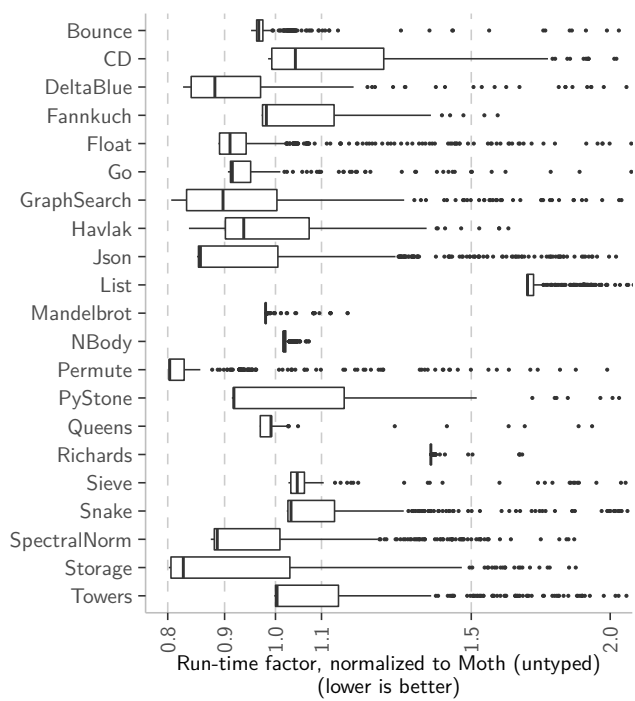

  \centering
	\TypingOverhead{}
	\caption{A boxplot comparing the performance of Moth
  with and without type checking.
  The plot depicts the run-time overhead on peak performance over
  the untyped performance. On average, transient type checking introduces
  an overhead of \OverheadTypingGMeanP (min. \OverheadTypingMinP, max. \OverheadTypingMaxP).
  The visible outliers reflect the noise in today's complex system
  and correspond \eg to garbage collection and cache effects.
  Note that the axis is logarithmic to avoid distorting the proportions
  of relative speedups and slowdowns.}
	\label{fig:typing-overhead}
\end{figure}

\cref{fig:typing-overhead} depicts
the overall results comparing Moth,
with all optimizations,
against the untyped version.
We see an average peak-performance overhead of 
\OverheadTypingGMeanP (min. \OverheadTypingMinP, max. \OverheadTypingMaxP).

The benchmark with the highest overhead of \OverheadListP is List.
The benchmark traverses a linked list and
has to check the list elements individually.
Unfortunately, the structure of this list introduces checks
that do not coincide with shape checks on the relevant objects.
We consider this benchmark a pathological case and discuss it
in detail in \cref{sec:disc-pathological-case}.

Beside List, the highest overheads are on
Richards (\OverheadRichardsP), CD (\OverheadCDP), 
Snake (\OverheadSnakeP), and Towers (\OverheadTowersP).
Richards has one major component, also a linked list traversal,
similar to List.
Snake and Towers primarily access arrays in a way that introduces checks
that do not coincide with behavior in the unchecked version.

In some benchmarks, however, the run time decreased; notably Permute (\OverheadPermuteP),
GraphSearch (\OverheadGraphSearchP), and Storage (\OverheadStorageP).
Permute simply creates the permutations of an array.
GraphSearch implements a page rank algorithm
and thus is primarily graph traversal.
Storage stresses the garbage collector by constructing a tree of arrays.
For these benchmarks the introduced checks seem to coincide with shape-check operations
already performed in the untyped version.
The performance improvement is possibly caused by having checks earlier,
which enables the compiler to more aggressively move them out of loops.
Another reason could simply be that the extra checks shift the boundaries
of compilation units.
In such cases, checks might not be eliminated completely,
but the shifted boundary between compilation units might mean that
the generated native code interacts better with
the instruction cache of the processor.

\paragraph*{Warmup Performance}

\begin{figure}[htb]
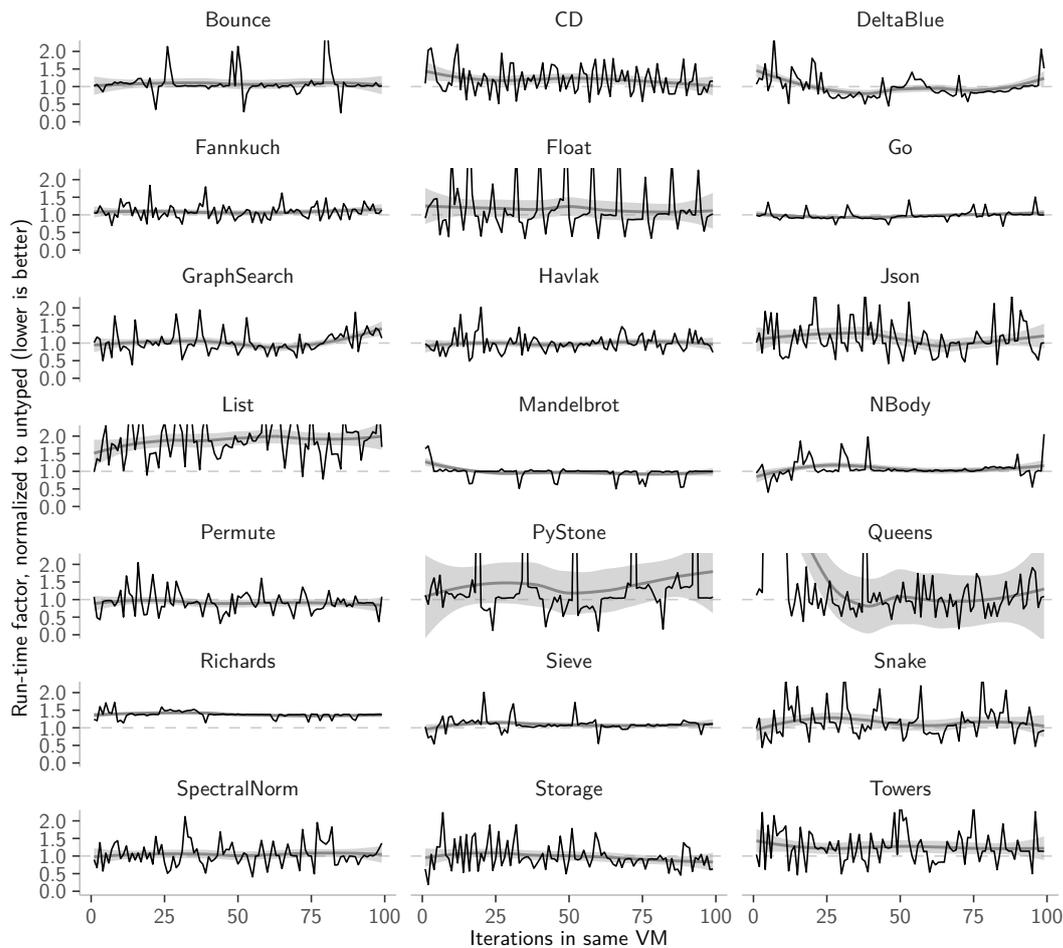

  \centering
	\TypingWarmup{}
	\caption{Plot of the run time for the first 100 iterations.
           The gray line is a local polynomial regression fit
           with a 0.95 confidence interval
           indicating the trend.
           The first iteration, \ie, mostly interpreted, seems
           to be affected significantly only for Mandelbrot.}
	\label{fig:typing-warmup}
\end{figure}

\Cref{fig:typing-warmup} shows the first 100 iterations for each benchmark.
The run-time factor is the result for the typed version over the untyped one.
Thus, any increase indicates a slow down because of typing.
The gray line indicates a smoothed version of the curve
based on local polynomial regression fitting\citep{Cleveland:1992}
using neighboring data points.
It also indicates a 0.95 confidence interval.

Looking only at the first iteration,
where we assume that most code is executed in the interpreter,
the overhead appears minimal.
Only the Mandelbrot benchmark shows a noticeable slowdown.
Benchmarks such as Float, PyStone, and Queens, however, show various spikes.
Since spikes appear in both directions (speedups and slowdowns),
we assume that they indicate a shift,
for instance, of garbage collection pauses,
which may happen because of different heap configurations
triggered by the additional data structures for type information.

\subsection{Effectiveness of Optimizations}

To characterize the concrete impact of our two optimizations, i.e.,
the optimized type checking node, which replaces complex type tests
with checks for object shapes, and our matrix to cache sub-typing information,
we look at the number of type checks performed by the benchmarks,
as well as the impact on peak performance.

\paragraph*{Impact on Performed Type Tests}

\Cref{tab:type-statistics} gives an overview of the number of type tests done
by the benchmarks during execution.
We distinguish two operations \code{check\_generic} and \code{is\_subtype\_of},
which correspond to the operations in
 \cref{ex:typenode:generic} and \cref{ex:type:satisfied} of \cref{ex:type}.
Thus, \code{check\_generic} is the test called
whenever a full type check has to be performed,
and \code{is\_subtype\_of} is the part of the check that
determines the relationship between two types.
The second column of \cref{tab:type-statistics} indicates
which optimization is applied,
and the following columns show the mean,
minimum, and maximum number of invocations of the tests
over all benchmarks.

\begin{table}[htb]
  \caption{Type Test Statistics over all Benchmarks.
  This table shows how many of the type tests can be avoided based on our two optimizations.
  With the use of an optimized node that replaces type checks with simple object shape checks,
  \code{check\_generic} is invoked only for the first time that a lexical location
  sees a specific object shape, which eliminates run-time type checks almost completely.
  Using our subtype matrix that caches type-check results,
  invocations of \code{is\_subtype\_of} are further reduced by an order of magnitude.}
  \label{tab:type-statistics}

  \begin{center}
    \TypingStatsTable{}  
  \end{center}
\end{table}

The baselines without optimizations are the rows with the results
for neither of the optimizations being enabled.
Depending on the benchmark,
we see that the type tests
are done tens of millions
to hundreds of millions times
for a single iteration of a benchmark.

Our optimizations reduce the number of type test invocations dramatically.
As a result, the full check for the subtyping relationship is done only once for 
any specific type and a possible super type.
Similarly, the generic type check is replaced by a shape check
and thus minimizes the number of expensive type checks
to the number of lexical locations that verify types
combined with the number of shapes a specific lexical
location sees at run time.

\paragraph*{Impact on Performance}

\begin{figure}[htb]
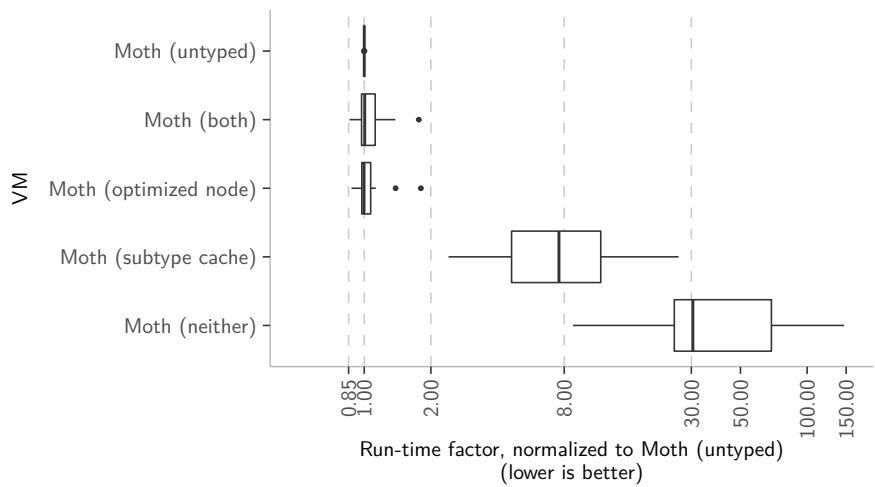

  \centering
	\OptimizationOverview{}
  \caption{Performance Impact of the Optimizations on the Average Peak Performance over all Benchmarks.
  The boxplot shows the performance of Moth normalized to the untyped version, \ie,
  without any type checks.
  The performance of Moth with both optimizations and Moth
  with only the node for optimized type checks are identical.
  The subtype check cache improves performance over the unoptimized version,
  but does not contribute to the peak performance.
  }
	\label{fig:perf-impact-optimization}
\end{figure}

\Cref{fig:perf-impact-optimization} shows how our optimizations contribute
to the peak performance.
The figure depicts Moth's average peak performance over
all benchmarks, depending on the activated optimizations.
As seen before, the untyped version is faster by \OverheadTypingGMeanP.
Moth with both optimizations enabled as well as
Moth with the optimized type-check node (cf. \cref{ex:type})
reach the same performance.
This indicates that the subtype cache matrix is not strictly necessary for
the peak performance.
However, we can see that the subtype cache matrix improves performance
by an order of magnitude over the Moth version without any type check optimizations.
This shows that it is a relevant and useful optimization.
Combined with the numbers of \cref{tab:type-statistics},
this optimization is going to be relevant for the very first execution of code.
A typical scenario of interest for developers would be, for instance, the
performance of unit tests, which has an major impact on developer productivity.

\subsection{Transient Typechecks are (Almost) Free}

As discussed in the introduction, in many existing gradually typed systems,
one would expect a linear increase of the performance overhead
with the increasing number of type annotations.

In this section, we show that this is not necessarily the case on our system.
For this purpose, we use two microbenchmarks Check and Nest,
which have at their core method calls with 5 parameters.
The Check benchmark calls the same method 10 times in a row, \ie, it has 10 call sites.
The Nest benchmark has 10 methods with identical signatures,
which recurse from the first one to the last one.
Thus, there are still 10 method calls, but they are nested in each other.
In both benchmarks, each method increments a counter,
which is checked at the end of the execution to verify that both do the same
number of method activations, and only the shape of the activation stack differs.

Each benchmark exists in 6 variants, going from no type annotations
over annotating only the first method parameter to annotating all 5 of the parameters.
Furthermore, we present the results for the first iteration as well as the hundredth iteration.
The first iteration is executed at least partially in the interpreter,
while the hundredth iteration executes compiled.

\begin{figure}
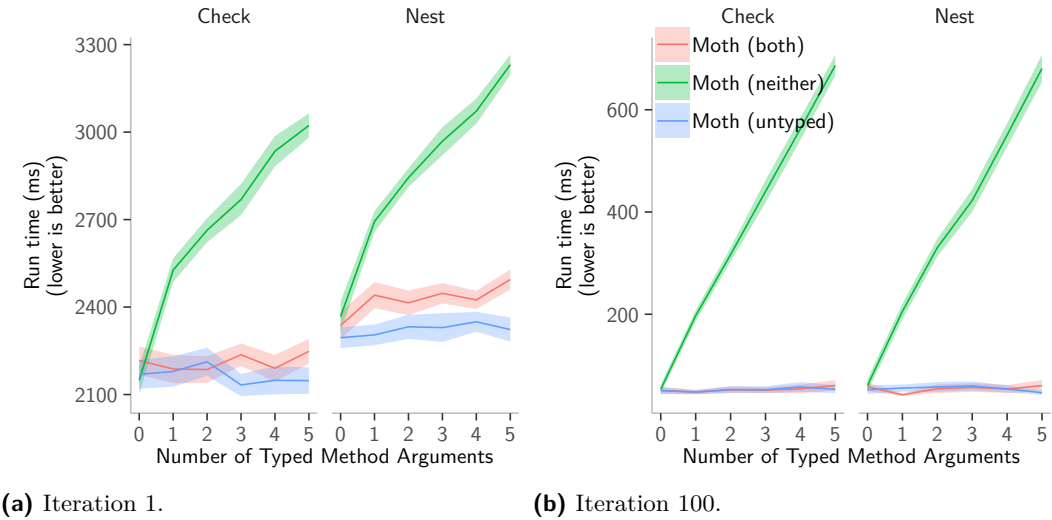

  \begin{subfigure}[t]{0.5\textwidth}
    \centering
    \TypeCostFirstIt{}
    \caption{Iteration 1.}
  \end{subfigure}\hfill
  \begin{subfigure}[t]{0.5\textwidth}
    \centering
    \TypeCostLastIt{}
    \caption{Iteration 100.}
  \end{subfigure}

  \caption{Transient Typechecks are (Almost) Free.
    Two microbenchmarks demonstrate the common scenario that the addition
    of type annotations in our system does not have an impact on peak performance.
    The two microbenchmarks are measured in 6 variants, stepwise increasing the number
    of method arguments that have type annotations.
    Furthermore, we show the result for the first benchmark iteration and the
    one hundredth.
    Moth (neither), \ie, Moth without our two optimizations sees a linear increase in run time.
    For the first iteration, we see some difference between Moth (both) and Moth (untyped).
    By the hundredth iteration, however, the compiler has eliminated
    the overhead of the type checks
    and both Moth variants essentially have the same performance
    (independent of the number of method arguments with type annotations).}
	\label{fig:type-cost-micro}
\end{figure}

\Cref{fig:type-cost-micro} shows that such a common scenario of methods being 
gradually annotated with types
does not incur an overhead on peak performance in our system.
The plot shows the mean of the run time for each benchmark configuration.
Furthermore, it indicates a band with the 95\% confidence interval.
The green line, Moth (neither), corresponds to our Moth with type checking
but without any optimizations.
For this case, we see that the performance overhead grows linearly.

For Moth (both) and Moth (untyped) we see for the first iteration that
the band of confidence intervals diverges, indicating that the additional type
checks have an impact on startup performance.
However, for the hundredth iteration, the confidence intervals overlap
for the optimized Moth as well as the one that does not perform typechecks.
This means that Moth does not suffer from
a general linear overhead for adding type checks.
Instead, most type checks do not have an impact on peak performance.
However, as previously argued for the List benchmark,
this is only the case for checks that can be subsumed by shape checks
(shape checks are performed whether or not type checks are presents).

\subsection{Changes to Moth}

Outlined earlier in \cref{sec:method}, a secondary
goal of our design was to enable the implementation of our approach to be
realized with few changes to the underlying interpreter.
This helps to ensure that each Grace implementation
can provide type checking in a uniform way.

By examining the history of changes maintained by our version control, 
we estimate that our implementation for Moth required
549 new lines and 59 changes to existing lines. 
The changes correspond to the implementation of 
new modules for the type class (179 lines) and 
the self-specializing type checking node (139 lines),
modifications to the front end to extract typing information
(115 new lines, 14 lines changes)
and finally the new fields and amended constructors for AST nodes 
(116 new lines, 45 lines changes).

\section{Discussion}
\label{sec:discussion}

\subsection{The VM Could Not Already Know That}
\label{sec:disc-pathological-case}

One of the key optimizations for our work
and the work of others\citep{Richards2017,Bauman2017}
is the use of object shapes
to encode information about types
(in our case),
or type casts and assumptions
(in the case of gradually typed systems).

The general idea is that a VM will already use object shapes
for method dispatches, field accesses, and other operations on objects.
Thus any further use to also imply type information
can often be optimized away
when the compiler sees that the same checks are done
(and therefore can be combined). This is similar to 
the elimination of other side-effect-free common subexpressions.

This assumption breaks, however, when checks are introduced
that do not correspond to those that exist already.
As described in \cref{sec:method},
our approach introduces checks for reading and writing to variables.
\Cref{ex:pathological-case} gives an example of a pathological case.
It is a loop traversing a linked list.
For this example our approach 
introduces a check, for the \code{ListElement} type,
when (1) assigning to and reading from \code{elem} and
(2) when activating the \code{next} method.
The checks for reading from \code{elem} and activating the method can be
combined with the dispatch's check on object shape.
Unfortunately, the compiler cannot remove the check
when writing to \code{elem}, because it has no information about
what value will be returned from \code{next}, and so it needs to preserve the check to be able to trigger an error
on the assignment.
For our List benchmark, this check induces an overhead of \OverheadListP.

\begin{lstlisting}[caption={Example for dynamic type checks not corresponding to existing checks.},escapechar=|,label={ex:pathological-case},float,floatplacement=htb]
var elem: ListElement := headOfList
while (...) do {
  elem := elem.next
}
\end{lstlisting}

\subsection{Optimizations}

As a simplification, we currently check variable access
on both reads and writes.
This approach simplifies the implementation, because we do not need to
adapt all built-ins, \ie,
all primitive operations provided by the interpreter.
One optimization could be to avoid read checks.
A type violation can normally only occur when writing to a variable,
but not when reading.
However, to maintain the semantics, this would require us to adapt
many primitives; 
such as operations that activate blocks to check their arguments,
or that write to variables or fields.
With our current implementation
we get errors as soon as user code accesses fields,
which simplifies the implementation.

Another optimization could be to use Truffle's approach to 
self-specialization\citep{Wurthinger:2012:SelfOptAST}
and propagate type information to avoid redundant checks.
At the moment, Truffle interpreters typically use self-specialization to 
specialize the AST to avoid boxing of primitive types.
This is done by speculating that some subtree always returns the expected type.
If this is not the case, the return value of the subtree is going to be
propagated via an exception, which is caught and triggers respecialization.
This idea could possibly be used to encode higher-level type information for
return values, too.
This could be used to remove redundant checks in the interpreter
by simply discovering at run time that whole subexpressions conform to the
type annotations.

\subsection{Threats to Validity}

This work is subject to many of the threats to validity common to
evaluations of experimental language implementations.  Our underlying
implementation may contain undetected bugs that affect the semantics
or performance of the gradual typing checks, affecting construct
validity --- we may not have implemented what we think we have. Given
that, our benchmarking harness run on the same implementation is
subject to the same risks, thus also affecting internal validity ---
we may not be measuring that implementation correctly.  Moth is built
on the Truffle and Graal toolchain, so we expect external validity
there at least --- we expect the results would transfer to other Graal
VMs doing similar AST-based optimizations.  We have less external
validity regarding other kinds of VMs (such as VMs specialized to 
particular languages, or VMs built via meta-tracing rather than partial evaluation). 
Nevertheless, we expect our results should be transferable
as we rely on common techniques.

Finally, because we are working in Grace, it is less obvious that our
results generalize to other gradually typed-languages. We have worked
to ensure e.g.\ our benchmarks do not depend on any features of Grace
that are not common in other gradually-typed object-oriented
languages, but as Grace lacks a large corpus of programs the
benchmarks are necessarily artificial, and it is not clear how the
results would transfer to the kinds of programs actually written in
practice. The advantage of Grace (and Moth) for this research is
that their relative simplicity means we have been able to build an
implementation that features competitive performance with significantly less
effort than would be required for larger and more complex languages.
On the other hand, more effort on optimisations could well lead to
even better performance.

\section{Related Work}
\label{sec:related-work}

Although syntaxes for type annotations in dynamic languages go back at
least as far as Lisp\citep{cltl2}, the first attempts at adding a
comprehensive static type system to a dynamically typed
language involved 
Smalltalk\citep{RalphJohnson1986}, with the first practical system
being Bracha's Strongtalk\citep{strongtalk}. Strongtalk
(independently replicated for Ruby\citep{DBRuby09}) provided a
powerful and flexible static type system, where crucially, the system
was \emph{optional} (also known as pluggable
\cite{GiladPluggable2004}). Programmers could run the static checker
over their Smalltalk code (or not); either way the type annotations
had no impact whatsoever of the semantics of the underlying Smalltalk
program.

Siek and Taha~\cite{Siek2006} introduced the term ``gradual typing''
 to describe the logical extension of this scheme: a
dynamic language with type annotations that could, if necessary, be
checked at runtime. Siek and Taha build on earlier complementary work extending fully statically typed languages with a ``\texttt{DYNAMIC}''
type---Abadi~\textit{et~al.}~'s 1991 TOPLAS paper~\cite{AbadiTOPLAS1991} is an
important early attempt
and also surveys previous work.

Revived practical adoption of dynamic languages generated revived
research interest, leading to the formulation of the ``gradual
guarantee''\citep{Siek2006,XXXSiek2015} to characterize sound gradual
type systems: removing type annotations should not change the
semantics of a correct program, drawing on Boyland's critical insight
that, of course, such a guarantee must by its nature forbid code that
can depend on the presence or absence of type declarations elsewhere
in the program\citep{Boyland2014}.
Because Moth only checks explicit type declarations (not inferred
intermediate types), Moth cannot not meet the refined gradual guarantee.
Moth ensures only that
the values passing through type annotations cannot be incompatible
with those annotations.

Type errors in gradual, or other dynamically checked, type systems will
often be triggered by the type declarations, but often those
declarations will not be at fault---indeed in a correctly typed
program in a sound gradually typed system,  the declarations cannot be
at fault because they will have passed the static type
checker. Rather, the underlying fault must be somewhere within the
barbarian dynamically typed code \emph{trans vallum}.
Blame tracking\citep{blame2009,blameThreesomes2010,blameForAll2011} localizes these
faults by identifying 
the point in the program where the system makes an 
assumption about dynamically typed objects, so can identify the root
cause should the assumption fail.  Different semantics for blame
detect these faults slightly differently, and impose more or less
implementation
overhead\citep{reticPython2014,monotonic2015,Vitousek2017}.

The diversity of semantics and language designs incorporating
gradual typing has been captured recently via surveys
incorporating formal models of different design options.
Chung et~al.\citep{kafka18} present an object-oriented model covering optional
semantics (erasure),  transient semantics, concrete semantics (from Thorn
\cite{Bloom2009}), and behavioural semantics (from Typed Racket), and
give a series of programs to clarify the semantics of a
particular language.  
Greenman et~al.\ take a more functional approach, again modelling
erasure, transient (``first order''), and behavioural (``higher
order'') semantics \cite{bensurvey18icfp}, and also present
performance information based on Typed Racket.
Wilson et~al.\ take a rather different approach, employing
questionnaires to investigate the semantics programmers expect of a
gradual typing system  \cite{shriramdls18}.

As with languages more generally, there seem to be two main implementation
strategies for languages mixing dynamic and static type checks: either
adding static checks into a dynamic language implementation, or adding
support for dynamic types to an implementation that depends on
static types for efficiency. Typed Racket, for example, optimizes code with
a combination of type inference and type declarations---the Racket
IDE ``optimizer coach'' goes as far as to suggest to programmers type
annotations that may improve their program's performance\citep{optimizerCoach2012}. In these implementations, values flowing
from dynamically to statically typed code must be checked at the
boundary.  Fully statically typed code needs no dynamic type checks,
and so generally performs better than dynamically typed code. Adopting
a gradual type system such as Typed Racket\citep{typedScheme08} allows
programmers to explicitly declare types that can be checked statically,
removing unnecessary overhead.

On the other hand, systems such as Reticulated Python\citep{reticPython2014}, SafeTypeScript\citep{Richards2017}, and our
work here, take the opposite approach.
These systems do not use information from type
declarations to optimize execution speed, rather the necessity to
perform (potentially repeated) dynamic type checks tends to slow
programs down, so here code with no type annotations generally
performs better than statically typed code, or rather, code with many
type annotations. In the limit, these kinds of systems may only ever
check types dynamically and may not involve a static type checker at
all. 

As gradual typing systems have come to wider attention, the question of their
implementation overheads has become more prominent.  
Takikawa~\textit{et~al.}~\cite{Takikawa2016} asked ``is sound gradual typing
dead?'' based on a systematic performance measurement on Typed Racket.
The key here is their evaluation method, where they constructed a
number of different permutations of typed and untyped code, and
evaluated performance along the spectrum.
Bauman~\textit{et~al.}~\cite{Bauman2017} replied to Takikawa~\textit{et~al.}'s study, in which they used Pycket\citep{Pycket2015} (a tracing JIT for Racket)
rather than the standard Racket VM,
but maintained full gradually-typed Racket semantics.
Bauman~\textit{et~al.} are able to demonstrate most benchmarks
with a slowdown of 2x on average over all configurations.
Note that this is not directly comparable to our system,
since typed modules do not need to do any checks at run time.
Typed Racket only needs to perform checks at boundaries between typed and untyped modules,
however, they use the same essential optimization technique that we apply,
using object shapes to encode information about gradual types.
Muehlboeck and Tate~\cite{Muehlboeck2017} also replied to
Takikawa~\textit{et~al.}, using a similar benchmarking method applied
to Nom, a language with features designed to make gradual types easier
to optimize, demonstrating speedups as more type information is added
to programs.  Their approach enables such type-driven optimizations,
but relies on a static analysis which can utilize the type
information, and the underlying types are nominal, rather than
structural.

Most recently, Kuhlenschmidt~\textit{et~al.}~\cite{Kuhlenschmidt:2018:preprint} employ an
ahead of time (\ie traditional, static) compiler for a custom
language called Grift and demonstrate good performance for code where more than
half of the program is annotated with types, and reasonable
performance for code without type annotations.

\label{reticRW}
Perhaps the closest to our approach are
Vitousek~\textit{et~al.}~\cite{reticPython2014} (incl. \citep{Vitousek2017,Greenman2018})
and Richards~\textit{et~al.}~\cite{Richards2017}.
Vitousek~\textit{et~al.}~describe dynamically checking transient types
for Reticulated Python (termed ``tag-type'' soundness by Greenman
and Migeed~\cite{Greenman2018}).
As with our work, Vitousek~\textit{et~al.}'s transient checks inspect
only the ``top-level'' type of an
object.
Reticulated Python undertakes these transient type checks at different
places to Moth.  Moth explicitly checks type anotations, while
Reticulated Python implicitly checks whenever values flow from dynamic
to static types.
We refrain from a direct performance comparison since
Reticulated Python is an interpreter without just-in-time compilation
and thus performance tradeoffs are different.

Richards~\textit{et~al.}~\cite{Richards2017} take a similar implementation
approach to our work, demonstrating that key mechanisms such as object shapes
used by a VM to optimize dynamic languages can be used to eliminate most of
the overhead of dynamic type checks.
Unlike our work, Richards
implement ``monotonic'' gradual typing with blame, rather than
the simpler transient checks, and do so on top of an adapted Higgs
VM.
The Higgs VM implements a baseline just-in-time compiler based on
basic-block versioning\citep{Chevalier-Boisvert:2016:ITS}.
In contrast, our implementation of dynamic checks
is built on top of the Truffle framework for the Graal VM, and reaches
performance approaching that of V8 (cf. \cref{sec:baseline-perf}).
The performance difference is of relevance here since any small constant factors
introduced into a VM with a lower baseline performance can remain hidden,
while they stand out more prominently on a faster baseline.

Overall, it is unclear whether our results confirm the ones
reported by Richards~\textit{et~al.}~\cite{Richards2017},
because our system is simpler.
It does not introduce the polymorphism
issues caused by accumulating cast information on object shapes,
which could be important for performance.
Considering that Richards~\textit{et~al.} report ca. 4\% overhead
on the classic Richards benchmark, while we see \OverheadRichardsP,
further work seems necessary to understand the performance implications of
their approach for a highly optimizing just-in-time compiler.

\section{Conclusion}
\label{sec:conclusion}

As gradually typed languages become more common,
and both static and dynamically typed languages are
extended with gradual features,
efficient techniques for gradual type checking become more important.
In this paper, we have demonstrated that optimizing virtual machines enable
transient gradual type checks with relatively little
overhead, and with only small modifications to an AST interpreter.
We evaluated this approach with Moth, an implementation of the Grace language
on top of Truffle and Graal.
In our implementation, types are structural and shallow: a type
specifies only the names of members provided by objects, and not
the types of their arguments and results.
These types are checked on access to variables,
when assigning to method parameters, and also on return values.
The information on types is encoded as part of an object's shape,
which means that shape checks already performed in an optimizing dynamic
language implementation can be used to check types, too. 
Being able to tie checks to the shapes in this way is critical for 
reducing the overhead of dynamic checking.

Using the \AWFY benchmarks as well as a collection of benchmarks from the
gradual typing literature, we find that our approach to dynamic type checking
introduces an overhead of 
\OverheadTypingGMeanP (min. \OverheadTypingMinP, max. \OverheadTypingMaxP)
on peak performance.
In addition to the results from further microbenchmarks,
we take this as a strong indication that transient gradual types do not
need to imply a linear overhead compared to untyped programs.
However, we also see that interpreter and startup performance is indeed
reduced by additional type annotations.

Since Moth reaches the performance of a
highly optimized JavaScript VM such as V8,
we believe that these results are a good indication
for the low peak-performance overhead of our approach.

In specific cases, the overhead is still significant and requires further
research to be practical. Thus, future research should investigate how the
number of gradual type checks can be reduced without causing
the type feedback to become too imprecise to be useful.
One approach might increase the necessary changes to a language implementation,
but avoid checking every variable read.
Another approach might further leverage Truffle's self-specialization
to propagate type requirements and avoid unnecessary checks.

Finally, we hope to apply our approach to other parts of the spectrum
of gradual typing, eventually reaching 
full structural types with
blame that support the gradual guarantee.  
This should let us verify that
Richards~\textit{et~al.}~\cite{Richards2017}'s results generalize to highly optimizing virtual
machines, or alternatively, show that other optimizations for precise
blame need to be investigated.

\bibliography{references}

\end{document}